\documentclass[journal]{new-aiaa}
\usepackage{textcomp}

\usepackage{graphicx}
\usepackage{amsmath}
\usepackage[version=4]{mhchem}
\usepackage{siunitx}
\usepackage{caption}
\usepackage{subcaption}
\usepackage{amsmath}

\usepackage{longtable,tabularx}
\title{A Method for Generating Closely Packed Orbital Shells and the Implication on Orbital Capacity}

\author{Miles Lifson\footnote{Graduate Student, Department of Aeronautics and Astronautics. Email: {\tt\small mlifson@mit.edu}, Student AIAA Member.}}
\affil{Massachusetts Institute of Technology, Cambridge, MA, 02139}
\author{David Arnas\footnote{Assistant Professor, Department of Aeronautics and Astronautics. Email: {\tt\small darnas@purdue.edu}.}}
\affil{Purdue University, 701 W. Stadium Ave., West Lafayette, IN, 47907}
\author{Martin Avendaño\footnote{Professor, PDI Funcionario, Department of Algebra, Geometry, and Topology. Email: {\tt\small mavend01@ucm.es}.}}
\affil{Universidad Complutense de Madrid, Plaza de Ciencias 3, Spain, 28040}
\author{Richard Linares\footnote{Boeing Assistant Professor, Department of Aeronautics and Astronautics. Email: {\tt\small linaresr@mit.edu}, Senior AIAA Member.}}
\affil{Massachusetts Institute of Technology, Cambridge, MA, 02139}

\begin{document}

\maketitle

\begin{abstract}

Shell-wise orbital slotting in Low Earth Orbit (LEO) can improve space safety, simplify space traffic coordination and management, and optimize orbital capacity. This paper describes two methods to generate 2D Lattice Flower Constellations (2D-LFCs) that are defined with respect to either an arbitrary degree or an arbitrary degree and order Earth geopotential.  By generating shells that are quasi-periodic and frozen with respect to the Earth geopotential, it is possible to safely stack shells with vertical separation distances smaller than the osculating variation in semi-major axis of each shell or a corresponding Keplerian 2D-LFC propagated under an aspherical geopotential. This helps mitigate the single inclination per shell requirement in prior work by admitting more shells for a given orbital volume while retaining self-safe phasing in each shell. These methods exploit previous work on the Time Distribution Constellation formulation and designs of closed 2D-LFCs under arbitrary Earth geopotentials using RGT orbits. Factors that influence the widths and shapes of these frozen shells are identified. Simplified formulas for estimating shell geometry and thickness are presented. It is shown that sequencing shells to group similar or ascending inclinations improves capacity versus arbitrary inclination ordering.

\end{abstract}

\section{Introduction}

Very large satellite constellations totaling 100,000 satellites or more have been proposed for Low Earth Orbit (LEO), including nearly 40,000 proposed to the Federal Communications Commission (FCC) last November alone.\footnote{ A tally of these applications is available at \url{https://www.cnbc.com/2021/11/05/space-companies-ask-fcc-to-approve-38000-broadband-satellites.html} and actual applications can be viewed within the FCC's International Bureau Filing System \url{https://www.fcc.gov/general/international-bureau-filing-system}.} The nation of Rwanda, on behalf of the company E-Space, recently submitted an International Telecommunications Union (ITU) filing for a single constellation consisting of more than 300,000 satellites across 27 shells.\footnote{ITU references to these filings can be found at \url{https://www.itu.int/ITU-R/space/asreceived/Publication/DisplayPublication/32322} and \url{https://www.itu.int/ITU-R/space/asreceived/Publication/DisplayPublication/32323}.} Not all of these constellations will make it to orbit and fewer still will be completed.  Nevertheless, this more than order-of-magnitude increase in the active satellite population will necessitate numerous technical, operational, and policy changes to ensure the long-term sustainability of the space environment in the face of denser operations.

\begin{figure*}[htbp]
    \centering
    \begin{subfigure}[b]{0.90 \textwidth}
      \includegraphics[width=1\linewidth]{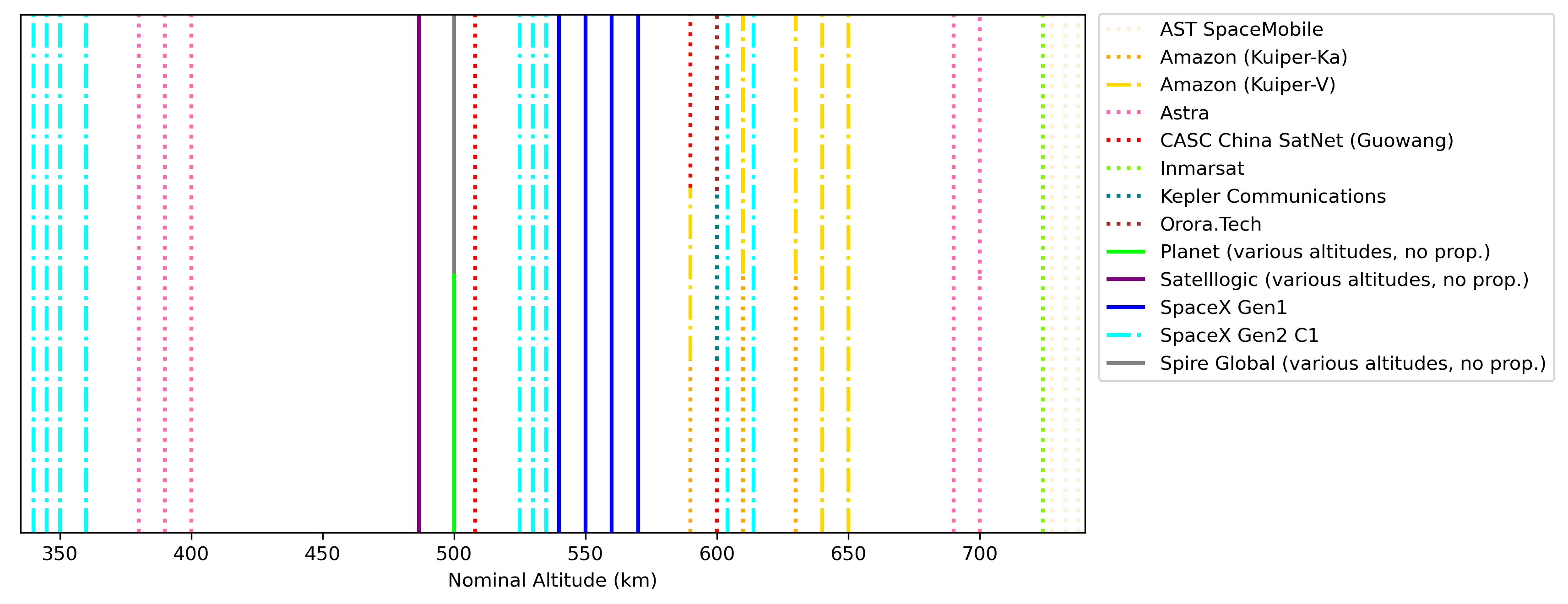}
      \caption{Selected Proposed Constellations >100 Satellites [335-740 km]}
    \end{subfigure}
    \begin{subfigure}[b]{0.90\textwidth}
      \includegraphics[width=1\linewidth]{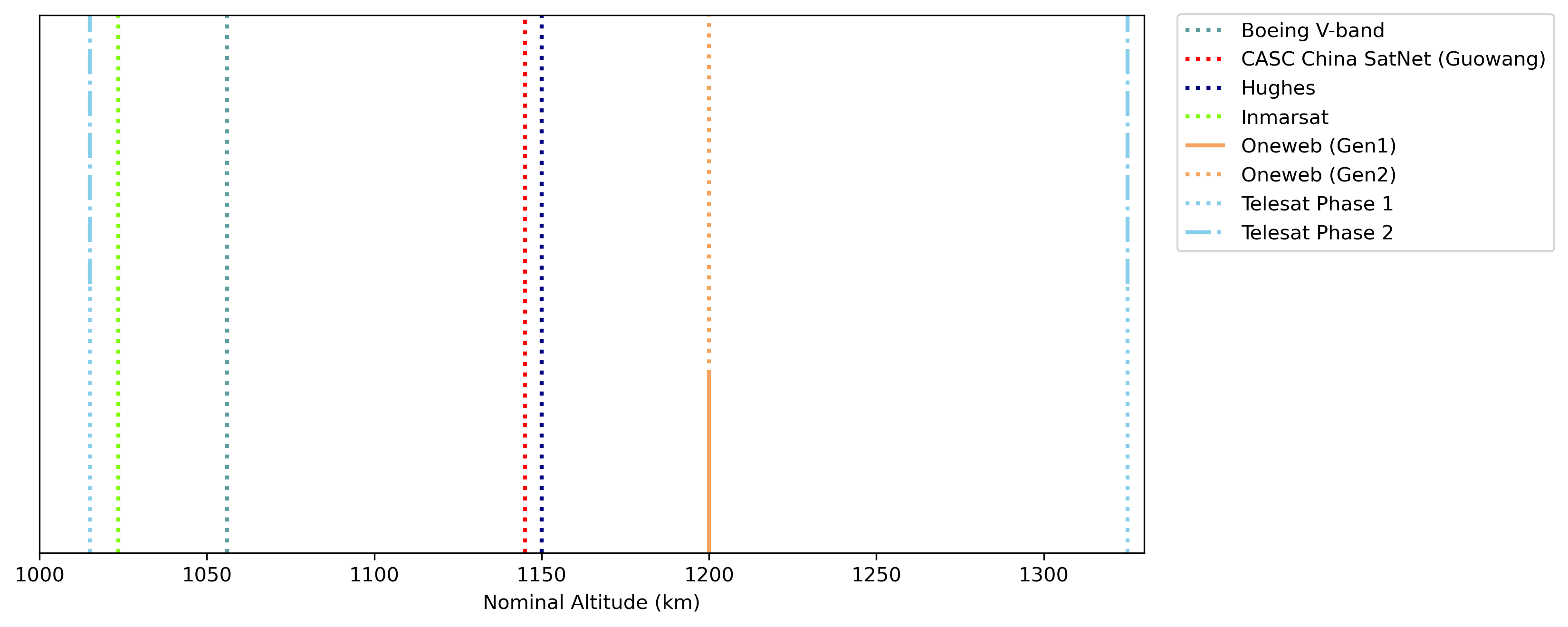}
      \caption{Selected Proposed Constellations >100 Satellites [1000-1330 km]}
    \end{subfigure}
\caption{There are significant overlaps and abutting large constellations planned within LEO.}
\label{fig:LEO} 
\end{figure*}

Figure \ref{fig:LEO} displays the nominal altitudes for selected proposed LEO constellations greater than 100 satellites. Worryingly, multiple large constellations have been proposed with limited orbital separation or overlapping nominal semi-major axes.\footnote{Because operator FCC filings claim large nominal orbital tolerances (often $\pm30$ km) to account for both operations and maneuvers, and because these filings frequently do not distinguish between mean and osculating semi-major axis or describe intended eccentricity vector maintenance information, it can be hard to understand actual intended nominal orbital variation for constellations based on FCC filings. Maintaining orbital shell separation accounting for the full range of orbital tolerances claimed by operators in FCC filings would be wasteful and prohibitively restrictive on orbital capacity. Even separation based simply on maximum and minimum osculating variation in semi-major axis would place significant limits on current and future capacity.} Large constellations with overlapping orbital altitude have the potential to significantly increase orbital conjunction risk and the operational burden associated with planning and coordinating collision avoidance maneuvers for active spacecraft controlled by different operators. Reference \cite{Alfano2020} found that, for the region from approximately 500-600km, the addition of planned constellations would result in significantly increased conjunction risk with a large fraction of that risk coming from active versus active cross-operator conjunction events. This finding may understate risk given the numerous additional constellations proposed subsequent to that study.  Furthermore, because collision avoidance is not perfect, overlapping large constellations increases residual collision risk and thus reduces long-term orbital capacity, limiting future constellation placement options if risk is not adequately mitigated.

As seen in Figure \ref{fig:LEO}, overlapping or closely-spaced cross-operator shell pairs include:
\begin{itemize}
\item China Aerospace Science and Technology Corporation (CASC) China SatNet Guowang and Amazon Kuiper, both at 590 km;
\item Orora.Tech, Kepler Communications, and CASC China SatNet Guowang at 600 km;
\item SpaceX Starlink at 604 km and 614 km and Amazon Kuiper at 610 km;
\item Telesat at 1015 km and Inmarsat at 1023.5 km;
\item Inmarsat at 724 km and AST SpaceMobile at 727.5, both of whom intend to operate small equatorial shells that could potentially be combined into a single equatorial shell; and
\item CASC China SatNet Guowang at 1145 km and Hughes at 1150 km.
\end{itemize}

This paper seeks to provide study of a potential solution to impose safe segregation between orbital shells while improving orbital capacity relative to these alternatives by relying on the use of frozen orbits and taking advantage of the approximate latitude-based altitude dependency that these shells exhibit. Many Earth-observing missions have used frozen orbits to minimize radial orbit variation, beginning with Seasat \cite{Cutting1978OrbitAF}, ERS-1 and ERS-2 \cite{Rosengren1999}, and TOPEX/Poseidon \cite{Frauenholz1998}, and continuing to more modern Earth observation missions that are required to maintain a near fixed radial distance to a ground site \cite[880]{Vallado2007}. Reference \cite{Bombardelli2021} suggested using frozen orbits and specifically optimizing for minimum radial distance variation as a function of latitude in order to avoid cross-shell collision risk. The paper computes these orbits numerically in the presence of zonal and tesseral harmonics, lunisolar third-body perturbations, solar radiation pressure, and atmospheric drag. The same frozen orbit technique was used to define satellite locations within a single shell in Reference \cite{Reiland2021} and shown to reduce intra-shell conjunction frequency. Nevertheless, these methods do not explicitly enforce periodicity or provide methods to evaluate orbital capacity, and neither paper discusses how to nest these frozen shells.

In this work, we aim to explicitly study the nesting of concentric frozen orbital shells and to generate these shells using methods that, subject to numerical error and quasi-stability of frozen orbits, enforce periodicity and connect the generated shells to previous theoretical results on 2D Lattice Flower Constellations (2D-LFC) \cite{2DLFCJ} that allow estimation of orbital capacity, reconfiguration, and optimization for minimum separation distances \cite{Arnas2020, Arnas2021, Avendano2021, Arnas2021a}. Rather than optimize individual satellite trajectories, we approach the problem by defining operational volumes, or slots, that evolve in a quasi-periodic manner and in which particular satellites can operate according to whatever control laws and internal positions they see fit. Accordingly, this paper incorporates only the effect of the Earth's geopotential\footnote{As noted in Reference \cite{tesisarnas}, it is in principle possible to incorporate third-body effects using the same approach used in this paper to incorporate tesseral effects, but the gains are small enough to not justify the additional computation and control requirements for most missions} to preserve the evergreen satellite-agnostic nature of slot designs within a shell and cross-plane quasi-periodicity which can be used to provides useful collision avoidance guarantees for 2D-LFCs. 

A consequence of this decision is that shell designs are not necessarily as compact as satellite-optimized trajectories that may fit within them, but better connect to existing 2D-LFC slotting work, can be computed in reasonable time periods without access to super computing resources (although such resources were used in this paper for propagating very large constellations, brute force conjunction checking, and accelerating shell definition) \footnote{Constellations with many orbital planes with few satellites per plane are particularly expensive to calculate with techniques that require significant new computation per-plane.}, and do not make assumptions about the drag area, coefficient of reflectivity, mass, or other general properties of the slotted satellites during shell design (although values do need to be assumed for post-hoc full-force modeling for stability analysis). We further explore several important implications for shell design. These include the usefulness of constellation visualization in the latitude-altitude plane as an analysis tool for ensuring non-conjunction between adjacent shells, the influence of shell inclination on possible shell tessellation, and an assessment of simplified shell geometry estimation methods. Last, the methods described in this paper are demonstrated for potentially overlapping or adjacent shells in LEO. These results and conclusions likely have relevance even if other methods are used to generate frozen shells.

Previous work on orbital slotting using 2D-LFCs assumed Keplerian orbits, significantly limiting the fidelity with which analysis of separation between concentric shells could be carried out \cite{Arnas2020}. In this work, slotting is demonstrated using 2D-LFCs defined to include the effects of higher order and degree terms of the Earth's gravity field. Specifically, two numerical orbit freezing methods are demonstrated to define 2D-LFCs that preserve quasi-periodicity in the presence of the Earth's geopotential. These method draw on previously published methods for generating perturbed flower constellations \cite{Arnas2020NominalPotential, Casanova2015, tesisarnas, time}. Reference \cite{tesisarnas} proposed a methodology to define orbits and satellite constellations in a set of relative trajectories that were closed under the effects of periodic orbital perturbations such as the Earth gravitational potential. Later, Reference \cite{Arnas2020NominalPotential} extended this result to obtain a set of invariants in the satellite distribution that were preserved under these periodic perturbations. Reference \cite{time} describes the Time Distribution Constellation formulation and describes how to convert a 2D-LFC defined in terms of Right Ascension of the Ascending Node (RAAN) and either mean or true anomaly to a Time Distribution Constellation defined relative to a seed trajectory using a ground track angle and time offset \cite{arnas20212dMissions}.

This work makes several new contributions. 
\begin{itemize}
\item First, it develops methods for designing compatible sets of 2D-LFC shells that minimize latitude-dependent altitude variation while maintaining self-safe shell structure. This contribution includes the first large-scale demonstration of these techniques to produce intra-shell self-safe configuration under arbitrary geopotentials \cite{Arnas2020NominalPotential}.  
\item Second, it proposes that, for maximum orbital density, shells should be designed to be frozen and structured to nest radius-latitude curves (generally through sequential or at least similar inclinations among adjacent shells). Nevertheless, the proposed shelling structure does work using non-frozen orbits but at the cost of increased control requirements.
\item Third, it assesses performance of several simplified models of shell shape and determines that inclusion of $J_2$ short period effects in these models is necessary for reasonable fidelity in analyzing the geometry of shell nesting. 
\end{itemize}

The organization of this paper is as follows. In Section \ref{Preliminaries}, several concepts used in this paper are briefly summarized including 2D-LFCs, classical orbital freezing techniques, equinoctial elements, and equations for several types of orbital perturbations. In Section \ref{Method}, two methods for frozen shell designs are described, one that accounts for only zonal terms in the Earth's geopotential and one that can account for both zonal and tesseral terms. Several models for simulating shell geometry without full shell computation are presented and compared. The latitude-altitude plane is demonstrated as an intuitive analysis tool to preserve non-conjunction between adjacent shells. In Section \ref{Results}, a demonstration is provided of gains to orbital density through sequential ordering of shell inclinations; a sample 2D-LFC is constructed using the zonal technique and demonstrated to maintain slot separation; and examples are presented for each of the two shell-design methods to address currently planned overlapping or adjacent orbital shells. 

\section{Preliminaries}\label{Preliminaries}

\subsection{2D Lattice Flower Constellations}
2D-LFCs \cite{2DLFCJ} are a special type of flower constellations that can represent all possible uniform distributions of satellites sharing values for same semi-major axis $a$, inclination $i$, eccentricity $e$, and argument of perigee $\omega$. In a 2D-LFC, admissible satellite slot locations are defined in terms of Right Ascension of the Ascending Node ($\Omega_{ij}$) and Mean Anomaly ($M_{ij}$) based on three integer values: the number of orbital planes ($N_o$), the number of satellites per orbit ($N_{so}$), and the configuration number ($N_{c}$), which acts as a phasing parameter between orbital planes according to the following equation:
\begin{equation}
    \label{f:02}
    \begin{bmatrix} N_o & 0\\ N_c & N_{so}\end{bmatrix} \begin{Bmatrix} \Omega_{ij}\\ M_{ij}\end{Bmatrix} = 2 \pi \begin{Bmatrix} i -1 \\ j - 1\end{Bmatrix}
\end{equation}
where $i \in [1, \cdots, N_o]$ and $j \in [1, \cdots, N_{so}]$ name the $j$-th satellite on the $i$-th orbital plane of the constellation. The configuration number is defined in the range $N_c \in [0, N_{o} - 1]$  to avoid redundant configurations. Critically, the periodicity of 2D-LFCs provides useful assurances for collision avoidance and shell capacity evaluation \cite{Arnas2020}.

\subsection{Non-spherical Earth Effects}

The Earth's geopotential is aspherical rather than perfectly uniform. These differences can be modeled using an infinite set of spherical harmonics in Equation \ref{geopotential}:
\begin{equation}
    \label{geopotential}
    U = \frac{\mu}{r} \left[ 1+ \sum_{l=2}^{\infty} \sum_{m=0}^{l} \left(\frac{R_\oplus}{r}\right)^lP_{l,m}[\sin(\phi_{gc_{sat}})]\{ C_{l,m} \cos(m \lambda_{sat}) + S_{l,m}\sin(m\lambda_{sat}) \} \right]
\end{equation}
where $\mu$ is the standard gravitational parameter of the Earth, $R_\oplus$ is the Earth's equatorial radius, $l$ is the degree of the geopotential and $m$ is the order, $P_{l,m}$ is the associated Legendre polynomial of degree $l$ and order $m$, $\phi_{gc_{sat}}$ is the geocentric latitude of the satellite, $\lambda_{sat}$ is geocentric longitude, and C and S are empirical constants provided by the chosen gravity field model. In practice, these are truncated based on the required fidelity in orbit propagation \cite[Ch. 8.6]{Vallado2007}. In this paper, the geopotential used to propagate a model is sometimes short-handed as $J_{l,m}$ where all zonal and tesseral terms up to degree $l$ and order $m$ are included. $J_l$ refers to either a specific zonal harmonic constants, $J_l = -C_{l,0}$, or a zonal-only model including all zonal terms up to degree $l$, depending on context. Taking the gradient of Equation \ref{geopotential} provides us with perturbing accelerations, which may then be numerically integrated using Cowell's formulation \cite[592]{Vallado2007}:
 \begin{equation}
    \label{a_geopotential}
    a_{nonspherical} = \nabla{\left( U-\frac{\mu}{r}\right)}
\end{equation}
The Orekit Astrodynamics library is used to perform much of the work in this paper. In Orekit, central body gravity forces are implemented based on the approach in Reference \cite{Holmes2002}.

\subsection{Classical Orbit Freezing}

Classical frozen orbit theory analysis considers only the $J_2$ and $J_3$ terms in the Earth's potential and seeks to find solutions to the Lagrange Planetary Equations that set the change to mean eccentricity, $\bar{e}$, and mean argument of perigee, $\bar{\omega}$, i.e. $\frac{d\bar{e}}{dt}$ and $\frac{d\bar{\omega}}{dt}$ to zero \cite{Vallado2007}. 
\begin{equation}
    \label{FrozenLPE_e}
    \frac{d\bar{e}}{dt} = -\frac{3}{2}\frac{\bar{n} }{(1-\bar{e}^2)^2}J_3 \left(\frac{R_\oplus}{\bar{a}}\right)^3 \sin{\bar{i}} \left(1-\frac{5}{4} \sin^2{\bar{i}}\right) \cos{\bar{\omega}}=0
\end{equation}
In this equation, $\bar{i}$ is the mean inclination and $\bar{n}$ is the mean mean motion.

\begin{equation}
    \label{FrozenLPE_omega}
    \frac{d\bar{\omega}}{dt} = \frac{3\bar{n}}{(1-\bar{e}^2)^2} J_2 \left( \frac{R_\oplus}{\bar{a}} \right)^2 \left( 1- \frac{5}{4} \sin^2{\bar{i}} \right) \left[ 1 + \frac{J_3}{2 J_2}\left( \frac{R_\oplus}{\bar{a}}\right) \frac{1} {(1-\bar{e}^2)} \left( \frac{\sin^2{\bar{i}}-\bar{e}^2\cos^2{\bar{i}}}{\sin{\bar{i}}} \right) \frac{\sin{\bar{\omega}}}{\bar{e}} \right]=0
\end{equation}

Equation \ref{FrozenLPE_omega} is satisfied for for $\bar{\omega} = \frac{\pi}{2}$ or $\bar{\omega} = \frac{3\pi}{2}$ radians (or at the critical inclination), meaning there will be no secular shift in mean argument of perigee. Equation \ref{FrozenLPE_e} implies there will be no secular shift in mean eccentricity when:
\begin{equation}
    \label{Frozen_e}
    \bar{e} = -\frac{1}{2}\frac{J_3}{J_2}\left( \frac{R_\oplus}{\bar{a}} \right) \left( \frac{1}{1-\bar{e}^2} \right) \left( \frac{\sin^2{\bar{i}}-\bar{e}\cos^2{\bar{i}}^2}{\sin{\bar{i}}} \right) \approx  -\frac{1}{2}\frac{J_3}{J_2}\left( \frac{R_\oplus}{\bar{a}} \right) \sin{\bar{i}}
\end{equation}

\subsection{Equinoctial Elements}

Equinoctial elements are a convenient way to represents near circular and near equatorial orbits, where some Keplerian elements suffer from ambiguity \cite{Broucke1972}. Given a set of Keplerian elements corresponding to a non-circular, non-equatorial orbit ($a, e, i, \omega, \Omega, v$), equinoctial elements can be expressed as (following the naming conventions of the Orekit astrodynamics library):
\begin{subequations} \label{equinoctialtoKep}
    \begin{align}
        a = a \\
        e_x = e \cos{\left(\omega + \Omega\right)} \\
        e_y = e \sin{\left(\omega + \Omega\right)} \\
        h_x = \tan\left({\frac{i}{2}}\right) \cos\left({\Omega}\right) \\
        h_y = \tan\left({\frac{i}{2}}\right) \sin\left({\Omega}\right) \\
        l_v = v + \omega + \Omega
    \end{align}
\end{subequations}

\subsection{Other perturbations}

This section briefly describes the modeling of other relevant orbital perturbations. Where described in detail in the Orekit documentation, this description matches their conventions. Where less information is provided, these descriptions follow Reference \cite{Vallado2007}. In the full force model in this work, atmospheric drag is assumed to be isotropic and modeled as: 
\begin{equation}
    \label{drag}
    \vec{a}_{drag} = -\frac{1}{2} \rho v^2 \frac{S}{m} \vec{D}\\
\end{equation}
where $\rho$ is the atmospheric density provided by the chosen atmospheric model, $v$ is the relative velocity between the spacecraft and the atmosphere, $m$ is the mass of the spacecraft,  $S$ is the product of the drag coefficient and cross-sectional area, and $\vec{D}$ is the drag coefficient vector. The value of $\rho$ is computed using DTM2000 \cite{Bruinsma2003}, a semi-empirical atmospheric model that depends on solar flux and geomagnetic data as well as atmospheric location and time. For this paper, solar and geomagnetic data was sourced from CelesTrak (\url{https://celestrak.com/SpaceData/}).
Third body accelerations are modeled according to:
\begin{equation}
    \label{3rdbody}
    \vec{a}_{3rd} = -\frac{\mu_\oplus \vec{r}_{\oplus sat}}{r^3_{\oplus sat}} + \mu_3 \left( \frac{\vec{r}_{sat3}}{r^3_{sat3}} - \frac{\vec{r}_{\oplus 3}}{r^3_{\oplus 3}}\right)\\
\end{equation}
where $\mu_\oplus$ is the Earth's gravitational constant, $\mu_3$ is the gravitational constant of the third body,  $\vec{r}_{\oplus sat}$ is a vector from the Earth to the satellite, $\vec{r}_{sat 3}$ is a vector from the satellite to the third body, and $\vec{r}_{\oplus 3}$ is a vector from the Earth to the 3rd body \cite[574]{Vallado2007}. Lunar, Earth, and solar positions are derived from JPL DE440 ephemerides.

Solar radiation pressure is modeled assuming a constant exposed area and coefficient of reflectivity using: 
\begin{equation}
    \label{srp}
    \vec{a}_{srp} = -\frac{p_{srp} c_R A_{\odot}}{m} \frac{\vec{r}_{sat \odot}}{\left\|\vec{r}_{sat \odot}\right\|}\\
\end{equation}
where $p_{srp}$ is the force of solar pressure per unit area, $c_R$ is the coefficient of reflectivity of the spacecraft,  $A_{\odot}$ is the sun-exposed area of the spacecraft, $m$ is the mass of the spacecraft, and  $\vec{r}_{sat \odot}$ is a vector from the satellite to the sun \cite[581]{Vallado2007}.

Solid tides are modeled including pole tides and third body effects from the Sun and Moon. Following Orekit defaults, 12 points are used for tides field interpolation, with a time-step of 600 seconds. Because the equations are lengthy and the effect is small, these equations are not reproduced here.

\section{Method}\label{Method}

\begin{figure}[htbp]
        \centering
        \begin{subfigure}[b]{0.45\textwidth}
          \includegraphics[width=1\linewidth]{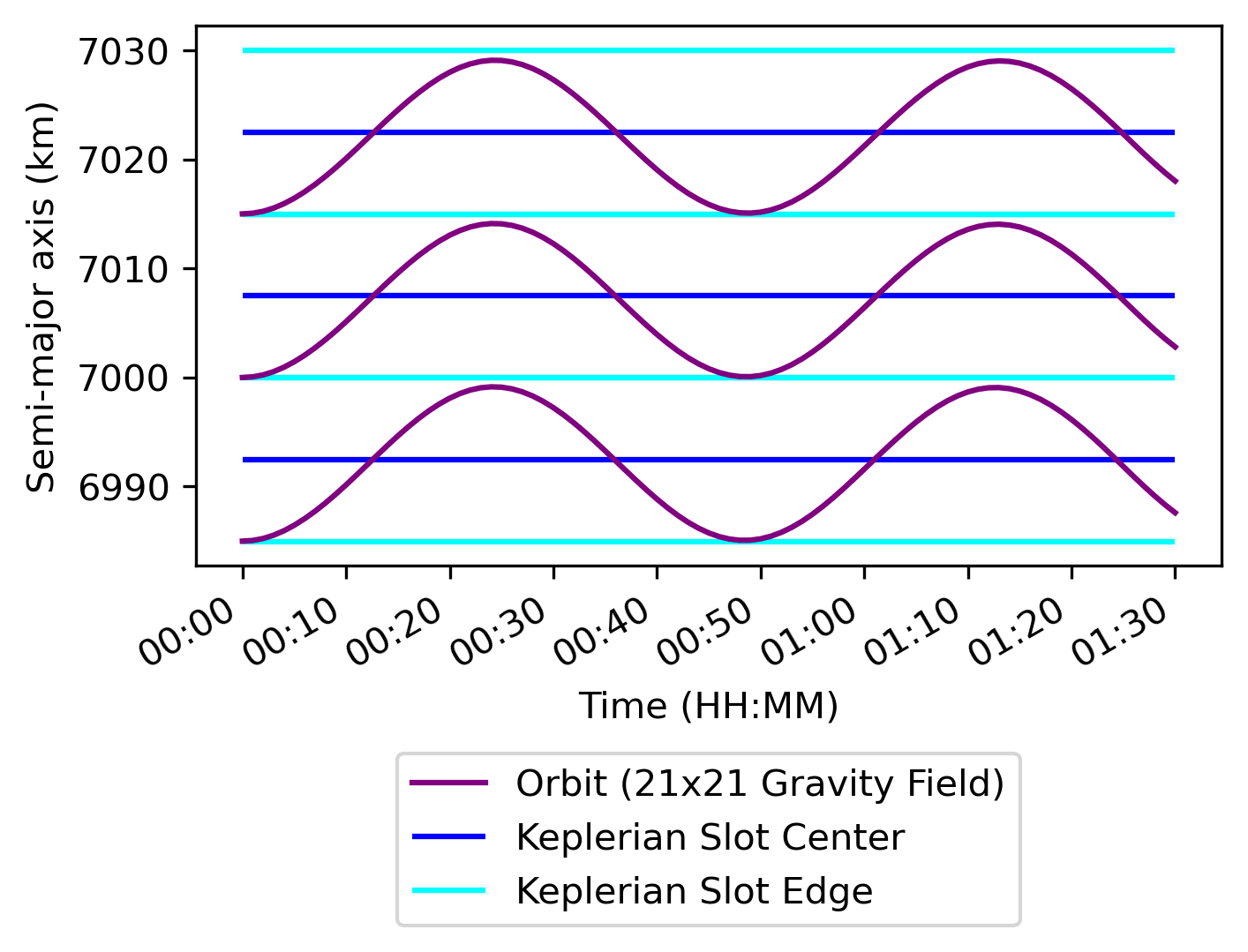}
          \caption{15 km Keplerian Slots (i = 60 degrees)}
        \end{subfigure}
        \begin{subfigure}[b]{0.45\textwidth}
          \includegraphics[width=1\linewidth]{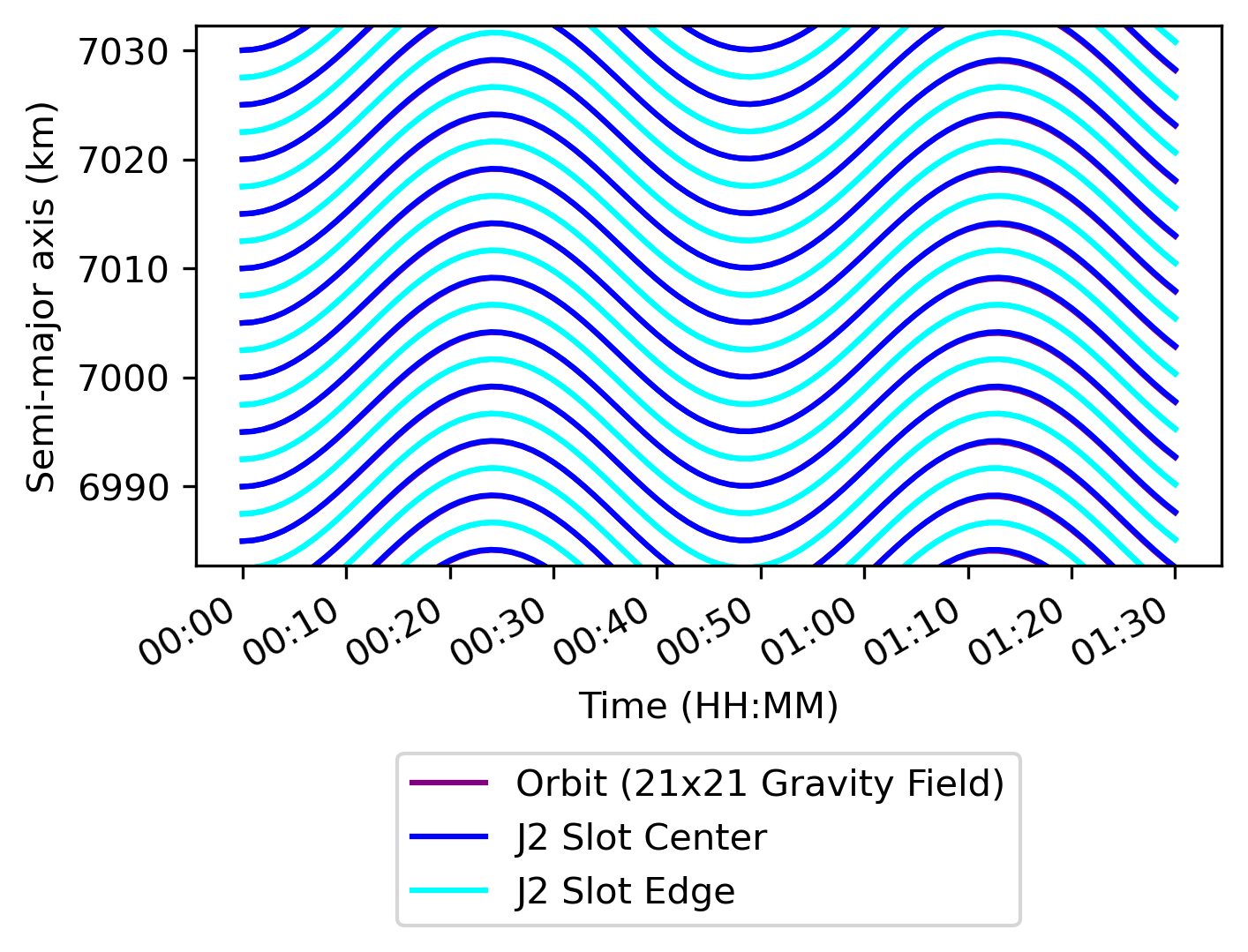}
          \caption{5 km Geopotential-Aware Slots (i = 60 degrees)}
          \label{fig:Slotsb} 
        \end{subfigure}
    \caption{Geopotential-aware slot altitude definition allows for much closer space of shells.}
    \label{fig:Slots} 
    \end{figure}

    In this section, two methods are described to define osculating 2D-LFCs in the presence of the Earth's geopotential. The first method assumes a zonal geopotential of arbitrary degree, while the second is applicable to a model with zonal, sectoral, and tesseral terms. In each, it is necessary to generate a numerically-closed seed orbit, which is used to define slot locations using a Time Distribution Constellation formulation \cite{time}. As seen in Figure \ref{fig:Slots}, the essential idea is that rather than defining orbital separation using fixed altitude bands, it is possible to define separation boundaries between shells accounting for the Earth's geopotential and enabling close shell stacking without generating overlaps in the latitude and altitude between adjacent shells. In Figure \ref{fig:Slotsb}, $J_2$ shells are defined, but any geopotential could be used. Note that the $J_2$ shell height is chosen for illustration and is not a recommendation or fundamental size limit. 
    
    Periodicity is a necessary condition to define a 2D-LFC. In the first method, periodicity is obtained over a single orbit through use of a genetic algorithm to find starting eccentricity conditions that minimize differences in the radial distance over the equator and shifts to the eccentricity vector. In the second method, periodicity is obtained over a repeating ground track (RGT) cycle for the orbit using the method developed in Reference \cite{Arnas2020NominalPotential}.
    
    \subsection{Definition of Osculating 2D-LFCs under a Zonal Geopotential} \label{geomethod}
    
    In this subsection, a 2D-LFC is mapped onto a zonal Earth geopotential of arbitrary degree. The approach exploits the fact that the gravity field is invariant with latitude and thus periodicity only requires radial closure and a frozen eccentricity vector.  This process has two steps. First, it is necessary to find a seed trajectory closed and frozen under the chosen zonal geopotential, $X_{0,0}$. Second, the 2D-LFC's RAAN and mean anomaly values can be mapped to ground track offset angles, $\Delta\Gamma_{i,j}$, and propagation times, $\Delta t{i,j}$, relative to $X_{0,0}$.
    
    Each frozen seed trajectory is found using a genetic algorithm (in this case, the differential evolution method within the SciPy library) with an objective function that seeks to preserve periodicity in the orbital radius and eccentricity vector over a single orbital period.\footnote{This approach was selected for simplicity and ease of implementation, although analytical solutions for frozen orbit eccentricity and argument of perigee exist for an arbitrary zonal geopotential model \cite{Rosborough1991}.} States are represented as initial osculating equinoctial orbits composed of $a$, $e_x$, $e_y$, $h_x$, $h_y$, and true longitude, $l_v$, with the algorithm free to change $e_x$ and $e_y$ within generous bounds $[lb_e, ub_e] = [-0.02, 0.02]$.
    \begin{equation}
    \begin{aligned}
        & \min{{\sqrt{\left(k_1|\vec{E}_f-\vec{E}_0|\right)^2 + \left(k_2\left(|r_f-r_0| + |r_f-r_m|\right)\right)^2}}} \quad \\
            &\quad \textrm{s.t.} \quad  e_x, e_y \in [lb_e, ub_e]  \\
            &\quad \quad \quad  \dot{\bf{X}} = f({\bf{X}(t)})  \\
    \end{aligned}
    \end{equation}
    where $\dot{\bf X} ={\bf f}(t,{\bf X}) $ is the dynamics in the orbital element space, $k_1 = \frac{1}{ub_e}$, and $k_2 = \frac{1}{500} \frac{1}{2} \frac{1}{a}$. For a given conservative disturbing function, in this case $R=U-\mu/r$, the Lagrangian planetary equations \cite[628]{Vallado2007} can be expressed in a non-singular form as:

    \begin{equation}
        \begin{aligned}
        \centering
        B &= \sqrt{1-e_y^2-e_x^2} \\
        C &= 1 + h_y^2 + h_x^2 \\
        \frac{d a}{dt} &= \frac{2}{n a} \frac{\partial R}{\partial l_v} \\
        \frac{d e_y}{dt} &= \frac{B}{n a^2}\frac{\partial R}{\partial e_x}-\frac{e_y B}{n a^2 (1+B)}\frac{\partial R}{\partial l_v} + \frac{e_x C}{2 n a^2B} \left\{ h_y \frac{\partial R}{\partial h_y} + h_x \frac{\partial R}{\partial h_x}\right\} \\
        \frac{de_x}{dt} &= -\frac{B}{n a^2} \frac{\partial R}{\partial e_y}-\frac{e_x B}{n a^2 (1+B)}\frac{\partial R}{\partial l_v} - \frac{e_y C}{2 n a^2B} \left\{ h_y \frac{\partial R}{\partial h_y} + h_x \frac{\partial R}{\partial h_x}\right\} \\
        \frac{d h_y}{dt} &= - \frac{h_y C}{2 n a^2 B} \left\{ e_x \frac{\partial R}{\partial e_y} - e_y \frac{\partial R}{\partial e_x} + \frac{\partial R}{\partial l_v} \right\} + \frac{f C^2}{4 n a^2} \frac{\partial R}{\partial h_x} \\
        \frac{d h_x}{dt} &= - \frac{h_x C}{2 n a^2 B} \left\{ e_x \frac{\partial R}{\partial e_y} - e_y \frac{\partial R}{\partial e_x} + \frac{\partial R}{\partial l_v} \right\} + \frac{f C^2}{4 n a^2} \frac{\partial R}{\partial h_y} \\        
        \frac{d l_v}{dt} &= n - \frac{2}{n a} \frac{\partial R}{\partial a} + \frac{B}{n a^2 (1+B)} \left\{ e_y \frac{\partial R}{\partial e_y} + e_x \frac{\partial R}{\partial e_x} \right\} + \frac{C}{2 n a^2 B} \left\{ h_y \frac{\partial R}{\partial h_y} + h_x \frac{\partial R}{\partial h_x} \right\} \\
        \end{aligned}
    \end{equation}

    This equation seeks to minimize the $\ell_2$ norm of the difference in the norm of the eccentricity vector (not to be confused with $e_x$ and $e_y$)  at the start and end of the orbit and the differences between the final equatorial orbital radius and the starting and half-orbit orbital radius. The values of $k_1$ and $k_2$ were chosen arbitrarily to yield good results in both high and low LEO by scaling the two components in a way that corresponds closely to minimizing overall orbital radius. If $ub_e$ is changed significantly, $k_1$ can be adjusted to maintain rough comparability in the magnitude of the two terms. The $\frac{1}{2}$ factor in $k_2$ is left to emphasize that two radius norms are being considered: the difference between the mid-orbit radial distance over the equator and final radius, and the difference between the beginning and final radial distance over the equator. Numerical simulations determined that in both low and high LEO, shell width across several examples was insensitive to moderate changes in the ratio between $k_1$ and $k_2$, but that extreme shifts (several orders of magnitude) could lead to thicker, less stable shells and/or extend the time for genetic algorithm convergence. To generate an optional starting guess, the mean $J_2$-$J_3$ frozen orbit elements are converted to osculating elements using the method from Kwok 1987 presented in Reference \cite{Vallado2007}, treating them as if they were produced with a $J_2$-only theory. This is not strictly true, but sufficient for the purpose of generating an initial approximate guess. An arbitrary 2D-LFC is then defined in terms of Keplerian elements using Equation \ref{f:02} to populate the $\Omega, M$ space.
    
    Once a seed orbit is defined, a rotation about the Earth's axis of rotation can be conducted for the first satellites in each orbital plane of the 2D-LFC. Under a zonal geopotential, this rotation does not change energy levels or orbital period and thus no adjustment to the rotated osculating state is needed. This simplifies the resulting Time Distribution Constellation. The orbital period of the numerically-closed seed orbit, $T_z$, can be found by propagating the seed orbit from one ascending equatorial crossing to the next (or another equivalent period). The propagation time $t_{i,j}$ for the $j$-th satellite in the $i$-th orbital plane from the starting epoch of $X_{0,0}$ can be found by exploiting proportionality between the fraction of orbital period under the zonal model and mean anomaly for the Keplerian 2D-LFC:
     \begin{equation}
         t = \frac{M_k}{2\pi}T_z 
     \end{equation}
    
    \subsection{Definition of Osculating 2D-LFCs under a Zonal and Tesseral Geopotential}
    
    Under a geopotential model that no longer only considers zonal terms, the process is more complicated as rotations change energy levels (meaning the rotated state may need slight adjustments to preserve a RGT for the same period) and a single orbital period is no longer sufficient to ensure periodicity. Instead, it is possible to use the process described in \cite{Arnas2020NominalPotential}, which exploits the fact that the Earth's geopotential is reasonably modeled as time invariant and thus trajectories closed in the Earth-centered Earth-fixed (ECEF) frame according to Equation \ref{f:04} have approximate periodicity. 
    \begin{equation}\label{f:04}
        T_c = N_p T_\Omega = N_d T_{\Omega G}
    \end{equation}
    A closed trajectory in the ECEF frame, i.e. a RGT trajectory, must satisfy Equation \ref{f:04} where $T_c$ is the cycle period of the closed trajectory, $N_p$ is the number of orbital revolutions per cycle, $N_d$ is the number of Earth sidereal days per cycle, $T_\Omega$ is the nodal period of the orbit, and $T_{\Omega G}$ is the nodal period of Greenwich \cite{Arnas2020NominalPotential}.
    
    The first step is to find a RGT at an altitude/semi-major axis within a user-specified tolerance of the user's desired altitude. $J_2$-adjusted semi-major axis values can be calculated from $N_d$ and $N_p$ using the method in Reference \cite[Sec. 7.1]{tesisarnas}. The mean $\bar{a}$ is obtained by solving Equation \ref{J2} using a Newton-Raphson solver, where $\omega_e$ is the Earth's angular velocity.
    \begin{equation}\label{J2}
       \bar{a}^{7/2}-k_1\bar{a}^2-k_2 = 0
    \end{equation}
    \begin{equation}\label{J2k1}
        k_1 = \frac{N_d}{N_p}  \frac{\sqrt{\mu}}{\omega_e}
    \end{equation}
    \begin{equation}\label{J2k2}
        k_2 = k_1 \frac{3J_2R_\oplus^2}{4(1-\bar{e}^2)^2}\left[ (2-3 \sin^2(\bar{i}) \sqrt{1-\bar{e}^2} + 4 - 5 \sin^2{\bar{i}}-2\frac{N_p}{N_d}\cos{\bar{i}} \right]
    \end{equation}
    The solver is initialized using the Keplerian semi-major axis:
    \begin{equation}\label{a0}
       \bar{a}_0 = \left[\left(\frac{N_d}{N_p}\right)^2 \frac{\mu}{\omega_e^2}\right]^{1/3}
    \end{equation}
    Each successive update can be calculated using this equation until $|\bar{a}_{j + 1 } - \bar{a}_j|< tol$ for a user-defined tolerance.
    \begin{equation}\label{j2it}
       \bar{a}_{j+1} = \bar{a}_j-\frac{\bar{a}_j^{7/2}-k_1\bar{a}_j^2-k_2}{\frac{7}{2}\bar{a}_j^{5/2}-2k_1\bar{a}_j}
    \end{equation}
    
\begin{figure}[htbp]
        \centering
        \includegraphics[width=0.7\textwidth]{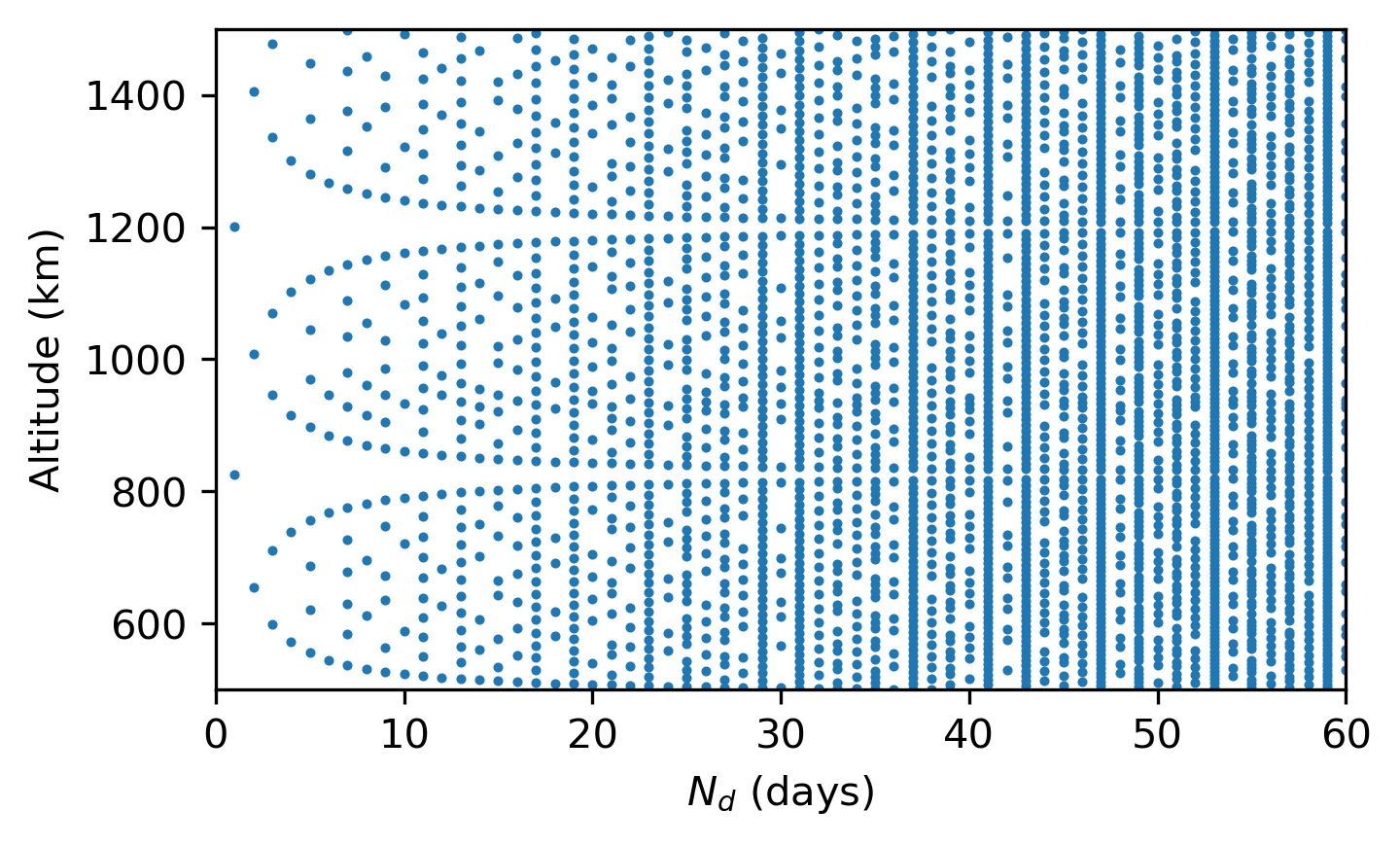}
        \caption{$J_2$-adjusted approximate admissible mean altitudes ($\mathbf{i = 45 \deg}$). Admissible altitudes can be made arbitrarily close by increasing $\mathbf{N_d}$ and finding corresponding $\mathbf{N_p}$.}
        \label{fig:admissible}
    \end{figure}

\begin{figure}[htbp]
        \centering
        \includegraphics[width=0.65\textwidth]{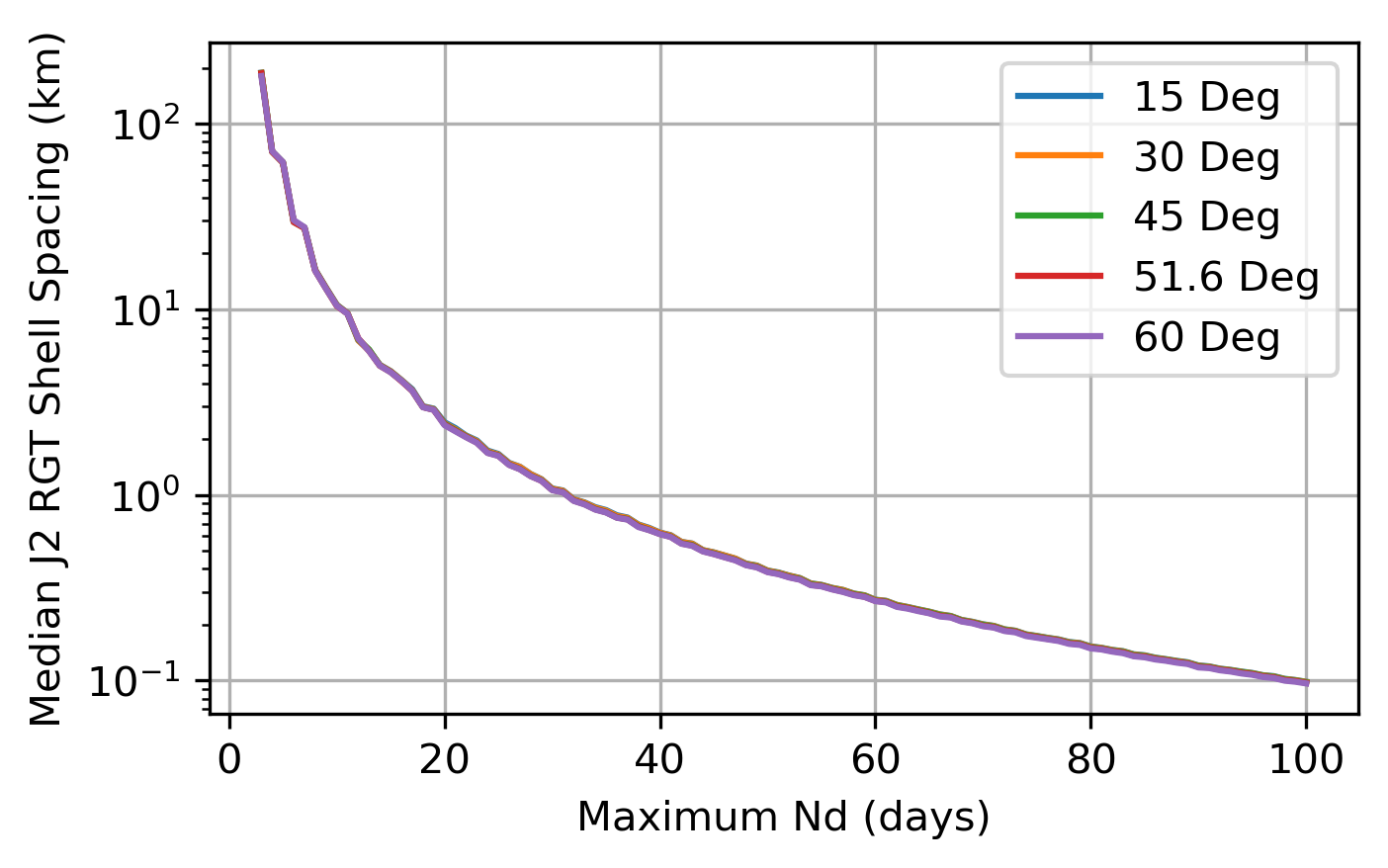}
        \caption{For an $\mathbf{N_d \leq 60}$ shells can be found within a median distance of about 250 meters from a given reference altitude.}
        \label{fig:rgtdists}
    \end{figure}

    Figure \ref{fig:admissible} shows altitudes for co-prime pairs of $N_d$ and $N_p$ that result in RGTs including $J_2$ for an inclination of $45\deg$. Changes in inclination slightly adjust the altitude of $N_d,N_p$ pairs due to changing the rate of $J_2$-induced RAAN precession, but do not impact the spacing of admissible altitudes, a result seen in the identical median shell spacing in Figure \ref{fig:rgtdists}. As demonstrated in these figures, the RGT condition is not particularly onerous. 

    Because the above method only closes orbits with respect to $J_2$, a further correction must be performed to close the reference trajectory in the presence of the full geopotential model. Using chosen $N_d$ and $N_p$ values, the seed trajectory is propagated from its initial state (transformed from mean to osculating elements using the method from Kwok 1987 presented in Reference \cite{Vallado2007}) for the full cycle period (obtained by counting the number of equatorial crossings) and the difference in initial and final longitude is used to find longitudinal drift. The initial osculating semi-major axis, $a$, is then iteratively modified using a secant-method solver to close the relative trajectory under the chosen gravity model by bringing the longitudinal drift arbitrarily close to zero. This process can be combined with the previous frozen search.
    
    Because rotations in the presence of tesserals potentially change energy levels, it is required to find a seed trajectory for each different ground-track in which the constellation is distributed. This means repeating the iterative process to find the conditions of semi-major axis that numerically close the ground-track for each individual ground-track of the space architecture. Once a seed trajectory is found, the actual cycle time, $T_c$,  can be calculated by propagating the seed relative trajectory for a full cycle. Reference~\cite{Arnas2020NominalPotential} derived a method to transform 2D-LFCs into time distribution constellations using the time and angle offsets in Equations \ref{eq:tdist} and \ref{eq:raandist}.
    \begin{equation}\label{eq:tdist}
        \Delta t_{ij} = \frac{T_c}{Np}\left[\frac{j-1}{N_{so}}-\frac{N_c (i-1)}{N_o N_{so}}\right] mod (T_c)
    \end{equation}
    \begin{equation}\label{eq:raandist}    
        \Delta \Gamma_{ij} = 2\pi\left[ \left(1-\frac{N_d}{N_p}\frac{N_c}{N_{so}}\right)\frac{i-1}{N_o} + \frac{N_d}{N_p}\frac{j-1}{N_{so}}\right] mod(2\pi)
    \end{equation}
    This formulation is able to determine both the distribution on a given ground-track, and also the different spacing between ground-tracks in the constellation. Once this transformation is performed, the seed trajectories obtained previously can be used to generate the initial positions of all the satellites of the constellation.
    
    \subsection{Simplified Shell Model}\label{ssec:simpleshellmodel}
    
    For optimization studies or initial compatibility checks, it can be desirable to understand the approximate shape of shells, without fully defining a 2D-LFC and propagating each satellite. This section describes and evaluates several simplified shell models suitable for these purposes. While bar notation is not used to avoid cluttering the equations, unless otherwise specified, all orbital elements in this and the next section are mean, rather than osculating elements.
    
    The classical orbital formula is expressed using the true anomaly, $\theta$, as a geometry-independent variable and a solution of Kepler's equation that relates $\theta$ to time \cite{Vallado2007}.
    \begin{equation}\label{range}    
    r(\theta) =\frac{a\left(1-e^2\right)}{1+e\cos\left(\theta\right)}
    \end{equation}
    Note that under perturbed motion, the radial distance curve diverges from Eq.~\ref{range}. Therefore we consider radius as an osculating quantity that changes with the osculating orbital elements. We can derive simplified expressions for the osculating radius using the mean elements as inputs.  Note that this equation is expressed in perifocal coordinates aligned with the eccentricity vector and therefore its relation to an inertial frame will also be a function of $i(t)$, $\omega(t)$, and $\Omega(t)$. 
    
    In general, radial distance is a function of both latitude and longitude, but for this work shell width is considered as a function of latitude only.
    Radial distance can be made a function of geocentric latitude angle, $\phi$, by noting that $\sin(\phi)=\sin(i)\sin(\theta+\omega)$. The radius of an osculating orbit can be expressed as a function of latitude angle using the following equation. 
    \begin{equation}\label{radius_lat}
    r(\phi)=\frac{a\,\left(1-e^2\right)}{1+e\,\cos\left(\omega-\mathrm{arcsin}\left(\frac{\sin\left(\phi \right)}{\sin\left(\mathrm{i}\right)}\right)\right)}
    \end{equation}
    Several important features are visible in this equation.  First, the variable $\omega$ has the effect of widening the curve of the $r(\phi)$ curve for values not equal to $\omega=\pi/2$ or $\omega=3\pi/2$. Additionally, $\omega=\pi/2$ and $\omega=3\pi/2$ produce curves that are identical after a reflection of $\phi$. Therefore, under this simplified model, selecting $\omega=\pi/2$ or $\omega=3\pi/2$ is preferred for maximizing orbital capacity (this condition is also used when designing $J_2/J_3$ frozen orbits). Equation \ref{radius_lat} can be simplified by assuming $\omega=\pi/2$:
    \begin{equation} \label{radius_lat_omega_fix}   
        r(\phi)=\frac{a\,\sin\left(\mathrm{i}\right)\,\left(1-e^2\right)}{\sin\left(\mathrm{i}\right)+e\,\sin\left(\phi \right)}=\frac{a\left(1-e^2\right)}{1+e\frac{\sin\left(\phi \right)}{\sin\left(\mathrm{i}\right)}}
    \end{equation}
    Second, note that when $\sin(\phi)$ is equal to $\sin(i)$ or $\sin(-i)$ the radial distance is $r(i)=a(1+e)=r_a$ and $r(-i)=a(1-e)=r_p$, respectively. Therefore, the curves have negative slope for $\omega=\pi/2$ and $0<i<\pi/2$ and positive slope for $\omega=3\pi/2$ and $0<i<\pi/2$. For best alignment, nearby shells should use the same argument of perigee to align the slopes of their shells.
    Third, it can be easily seen that the average slope of the radius vs latitude curve is given by $a e/i$. Non-zero eccentricity create sloped radius vs. latitude curves where larger semi-major axis values increase the slope.

    For greater fidelity, we can avoid assuming the value of $\omega$ and add the short period effect of $J_2$ on orbital radius by manipulating the following equation from \cite[709]{Vallado2007}, introduced by Reference \cite{Kwok1987}: 
    \begin{equation}\label{drsp}
    \Delta r_{sp} = -\frac{J_2 R^2_\oplus}{4p}\left( (3 \cos^2{(i)}-1) \left\{ \frac{2\sqrt{1-e^2}}{(1+e\cos{(\theta)})^2} + \frac{ e \cos{(\theta)}}{1+\sqrt{1-e^2}} + 1\right\} - \sin^2{(i)}\cos{(2u)} \right)
    \end{equation}
    where $p$ is the semiparameter and $u$ is the argument of latitude. The following substitutions can then be made:
    \begin{equation}\label{drspsub}
    \begin{gathered}
    p = a (1-e^2) \\
    u = \arcsin{\left(\frac{\sin{(\phi)}}{\sin{(i)}}\right)} \\
    \theta = \arcsin{\left(\frac{\sin{(\phi)}}{\sin{(i)}}\right)} - \omega
    \end{gathered}
    \end{equation}
     to yield:
    \begin{gather} \label{sub}
    \Delta r_{sp} = -\frac{J_2 R^2_\oplus}{4a (1-e^2)}\Bigg( (3 \cos^2{(i)}-1) \left\{ \frac{2\sqrt{1-e^2}}{\left(1+e\cos{\left(\arcsin{\left(\frac{\sin{(\phi)}}{\sin{(i)}}\right)} - \omega\right)}\right)^2} + \frac{ e \cos{\left(\arcsin{\left(\frac{\sin{(\phi)}}{\sin{(i)}}\right)} - \omega\right)}}{1+\sqrt{1-e^2}} + 1\right\} \nonumber \\
    \quad - \sin^2{(i)}\cos{\left(2\arcsin{\left(\frac{\sin{(\phi)}}{\sin{(i)}}\right)}\right)} \Bigg)
    \end{gather}
    and a modified osculating radius formula:
    \begin{equation}\label{radius_lat_drsp}
    r(\phi)=\frac{a\,\left(1-e^2\right)}{1+e\,\cos\left(\omega-\mathrm{arcsin}\left(\frac{\sin\left(\phi \right)}{\sin\left(\mathrm{i}\right)}\right)\right)} + \Delta r_{sp}
    \end{equation}
    To assess these various formulas for accuracy, equations \ref{radius_lat}, \ref{radius_lat_omega_fix}, and \ref{radius_lat_drsp} were implemented. While it was not necessary for this work, $\omega$ is potentially ambiguously defined for small eccentricities.  This same equation can be recast into an alternative formulation using the substitutions: $e_x = e \cos{(\omega)}$, $e_y = e \sin{(\omega)}$, $e = \sqrt{e_x^2 + e_y^2}$, and $cos(v) = \frac{e_x}{\sqrt{1+e \cos{(v)}}}\cos{(u)} + \frac{e_y}{\sqrt{1+e \cos{(v)}}}\sin{(u)}$, and the $u$ expression from Equation \ref{drspsub} to avoid this issue.
    
\begin{figure}[htbp]
        \centering
        \begin{subfigure}[t]{.45\textwidth}
            \centering
            \includegraphics[width=0.9\textwidth]{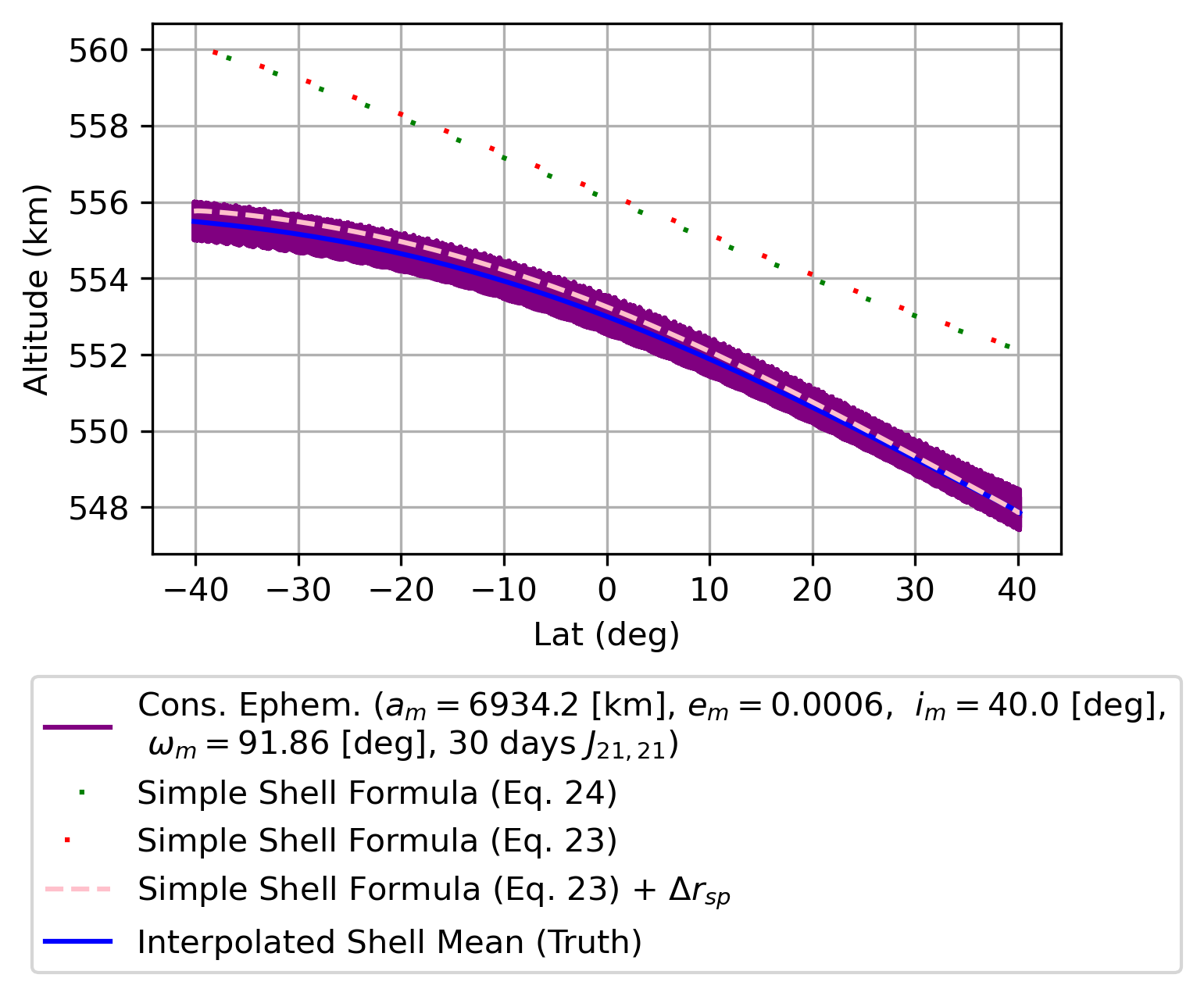}
            \caption{Equation Fit}
        \end{subfigure}
        \begin{subfigure}[t]{.45\textwidth}
            \centering
            \includegraphics[width=0.9\textwidth]{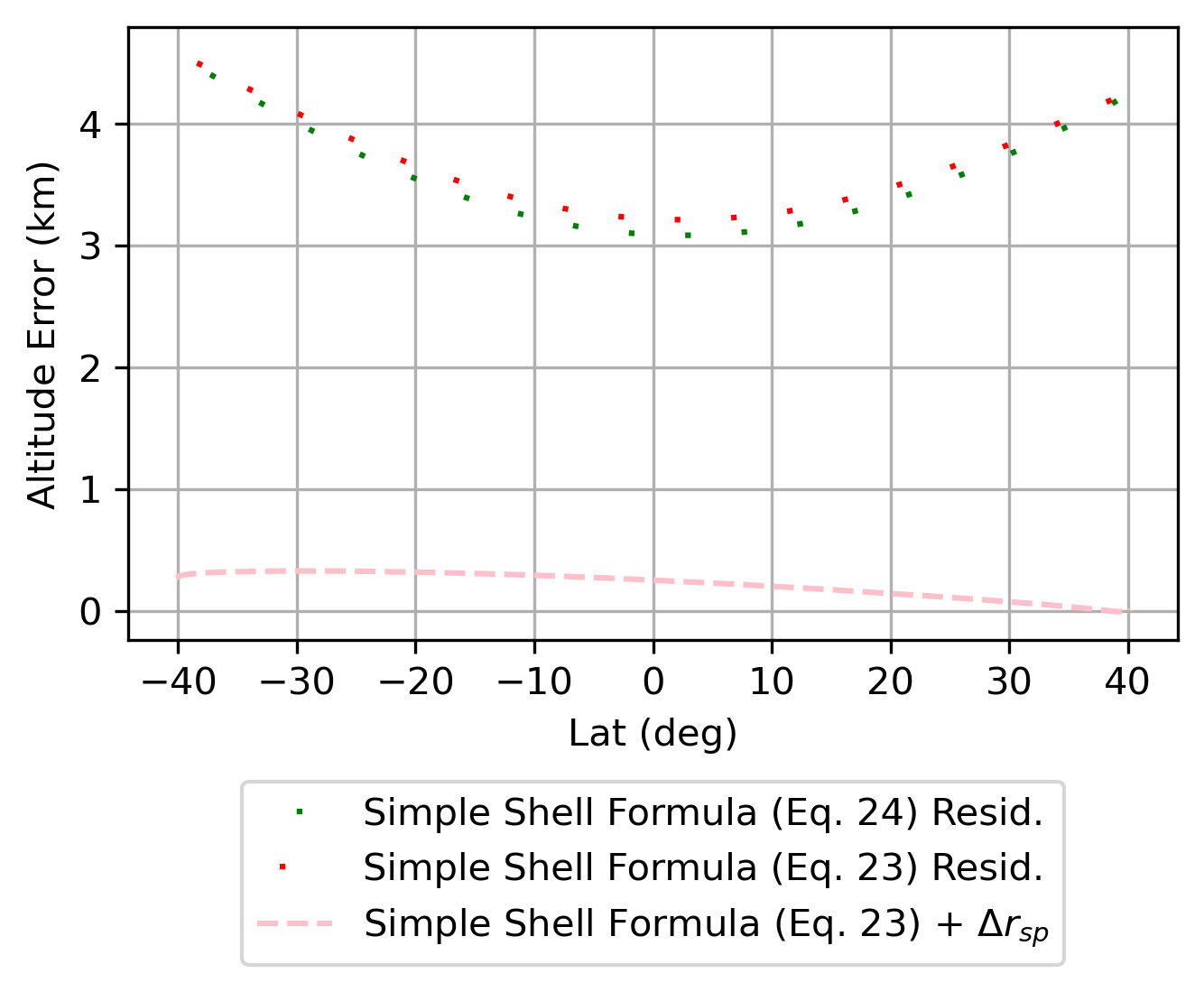}
              \caption{Fit Residuals}
        \end{subfigure}
        \caption{Mean to Osculating Shell Geometry Comparison Example 1 ($\mathbf{i = 40 [deg], N_o = 5, N_{so} = 5, N_c = 2}$)}
    \label{fig:inc40fit}
    \end{figure}

\begin{figure}[htbp]
        \centering
        \begin{subfigure}[t]{.45\textwidth}
            \centering
            \includegraphics[width=0.9\textwidth]{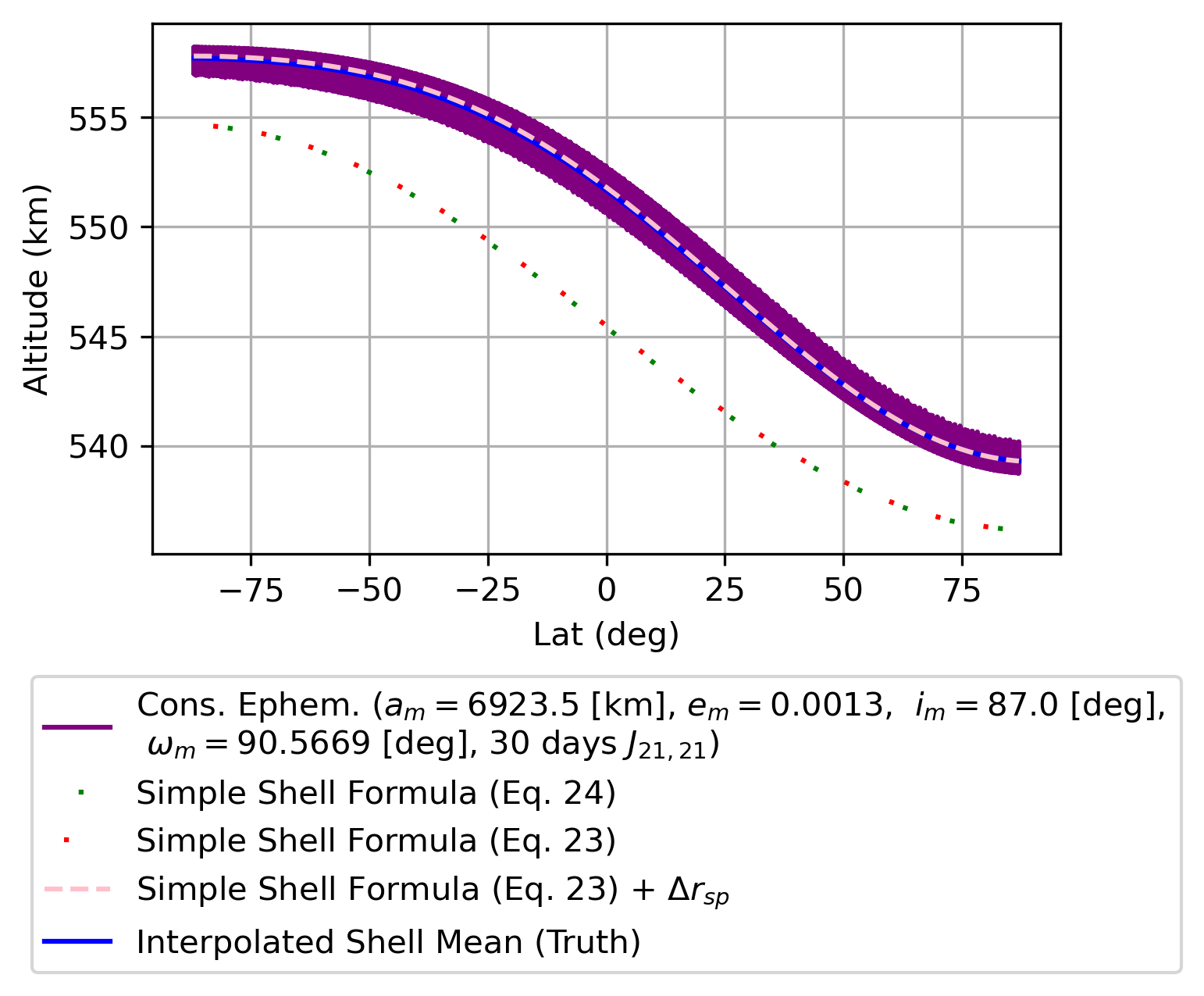}
            \caption{Equation Fit}
        \end{subfigure}
        \begin{subfigure}[t]{.45\textwidth}
            \centering
            \includegraphics[width=0.9\textwidth]{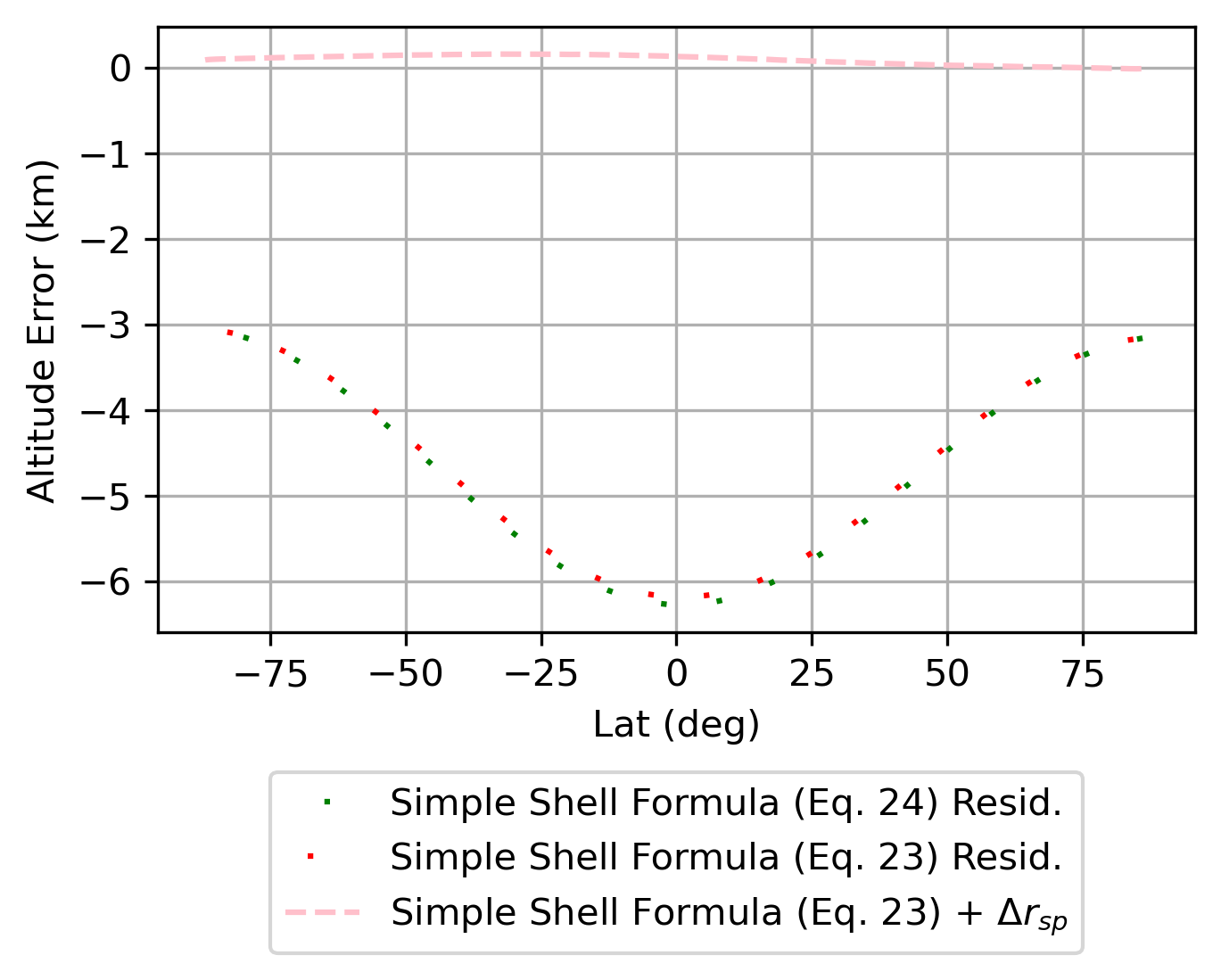}
              \caption{Fit Residuals}
        \end{subfigure}
        \caption{Mean to Osculating Shell Geometry Comparison Example 2 ($\mathbf{i = 87 [deg], N_o = 5, N_{so} = 5, N_c = 2}$)}
    \label{fig:inc87fit}
    \end{figure}

    These equations were then evaluated for goodness of fit using the five constellations in Figure \ref{fig:shellstacking}, although only two are shown in this paper for conciseness. Mean element values were recovered from the initial osculating state of the seed trajectory using Orekit's implementation of the Draper Semi-analytical Satellite Theory (DSST) for a model featuring only a $J_{21, 21}$ gravity model (mean elements as defined by DSST's theory are different than the $J_2$-only theory and method described earlier)\cite{bernard2015validating}. These recovered mean values were used as inputs into each of the three equations. For each constellation, the osculating altitude-latitude relationships for all satellites were plotted and an interpolated function was generated for the mean of these curves using a 90-point interpolated quadratic spline. As seen in Figures \ref{fig:inc40fit} and \ref{fig:inc87fit}, Equations \ref{radius_lat} and \ref{radius_lat_omega_fix} generate very similar results, with residuals of several kilometers and significant errors in shape, especially at more extreme latitudes.
    
\begin{figure}[htbp]
        \centering
        \begin{subfigure}[t]{.45\textwidth}
            \centering
            \includegraphics[width=0.9\textwidth]{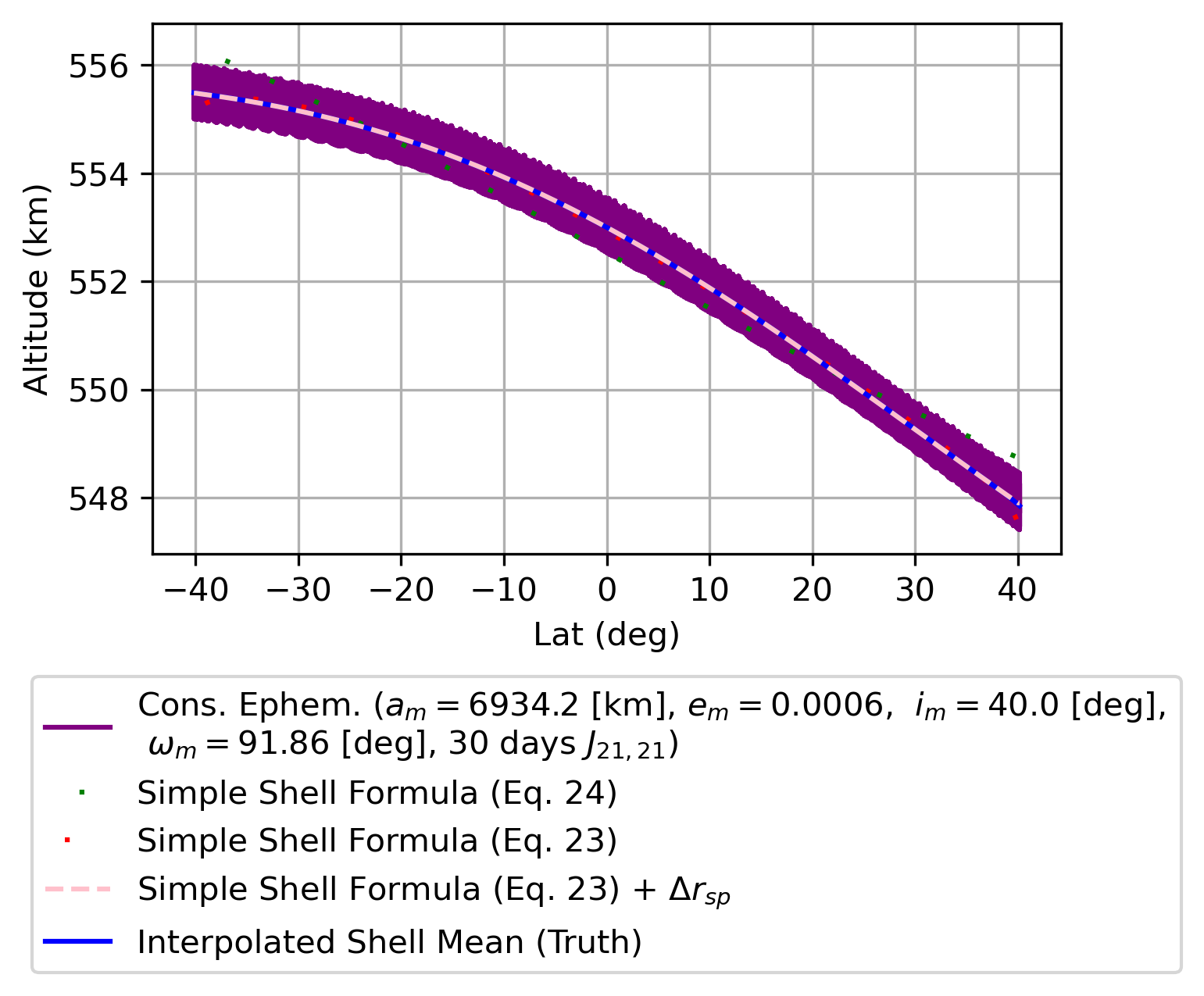}
            \caption{Equation Fit}
        \end{subfigure}
        \begin{subfigure}[t]{.45\textwidth}
            \centering
            \includegraphics[width=0.9\textwidth]{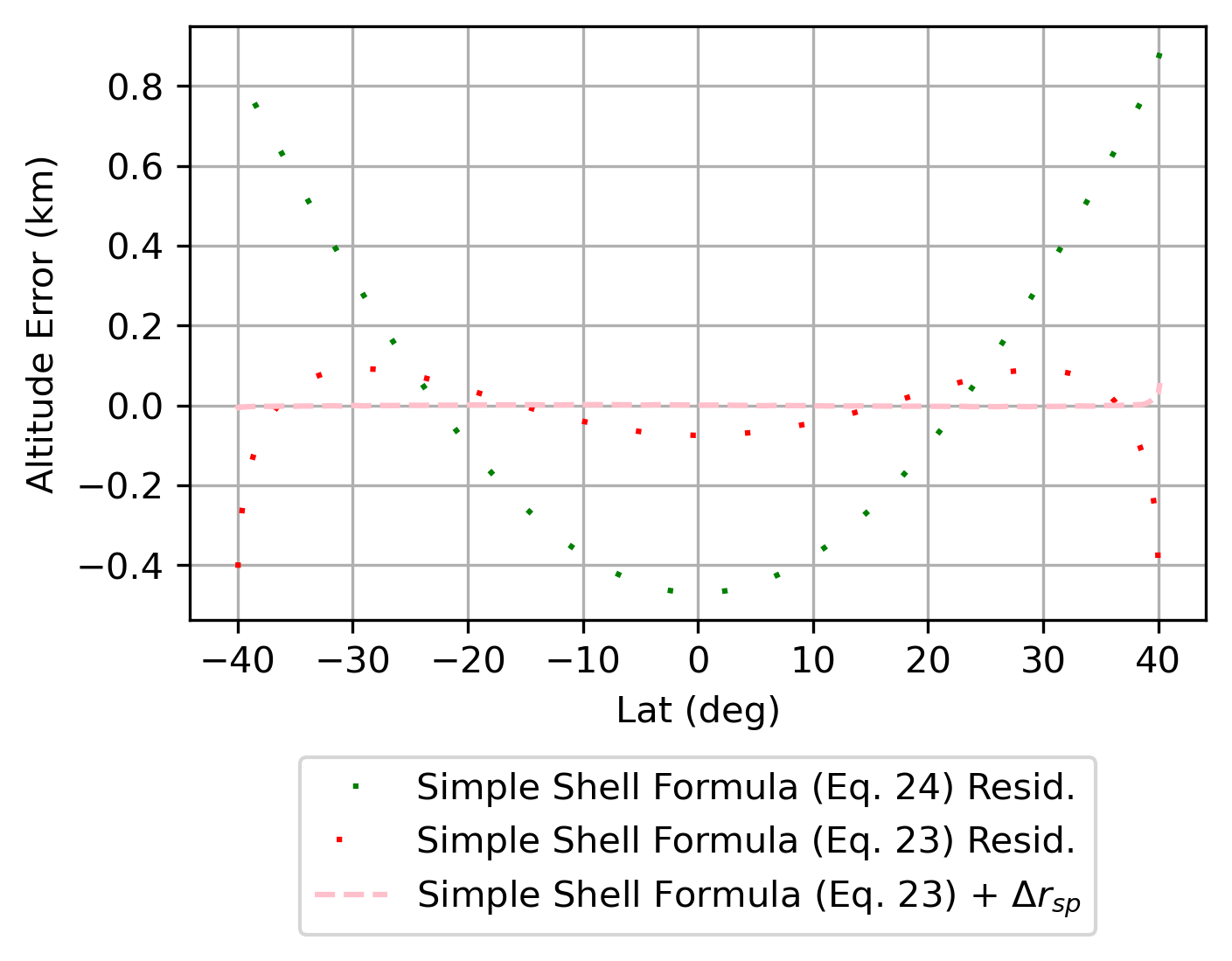}
              \caption{Fit Residuals}
        \end{subfigure}
        \caption{Genetic Algorithm Geometry Comparison Example 1 ($\mathbf{i = 40 [deg], N_o = 5, N_{so} = 5, N_c = 2}$)}
    \label{fig:inc40fitga}
    \end{figure}

\begin{figure}[htbp]
        \centering
        \begin{subfigure}[t]{.45\textwidth}
            \centering
            \includegraphics[width=0.9\textwidth]{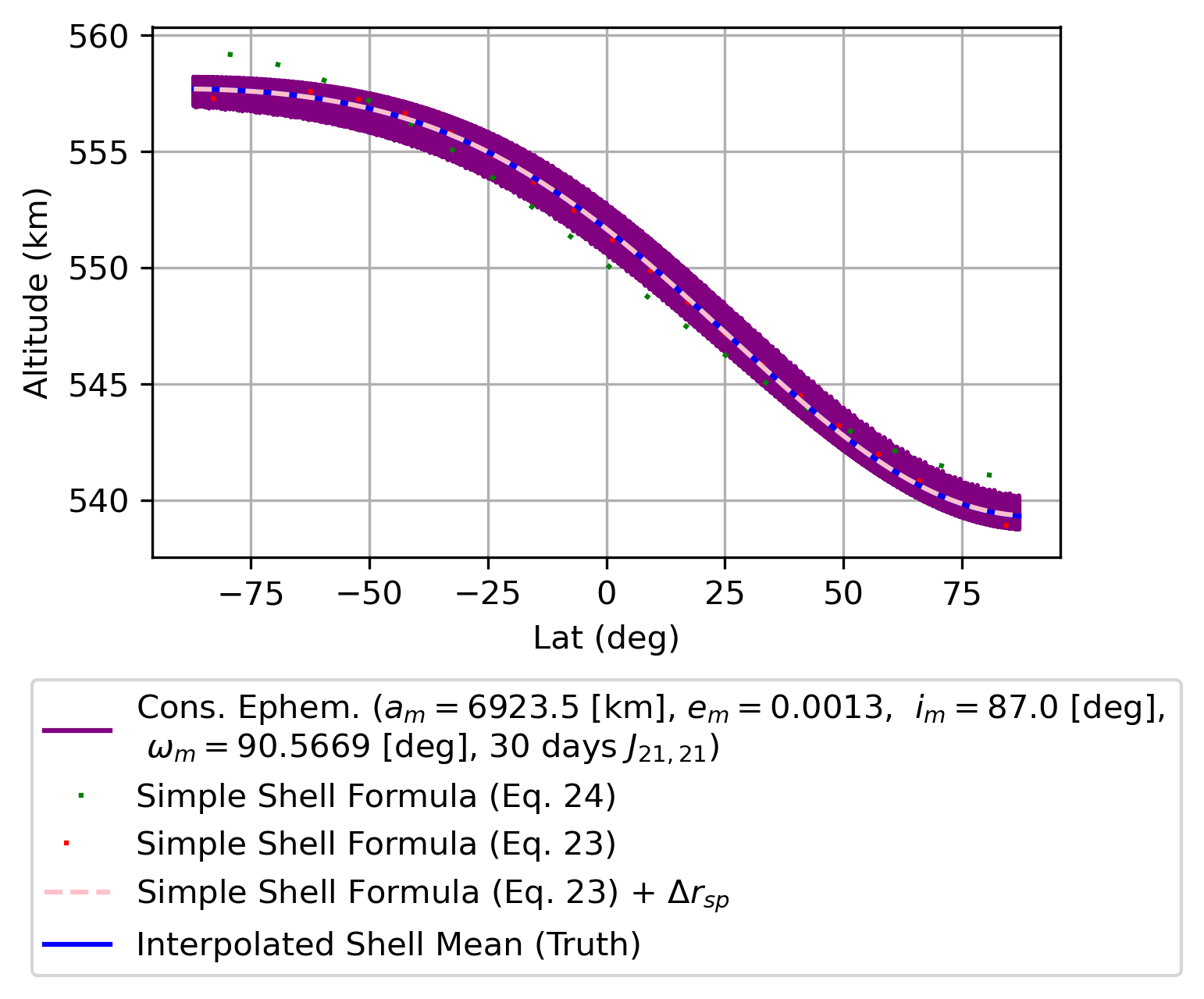}
            \caption{Equation Fit}
        \end{subfigure}
        \begin{subfigure}[t]{.45\textwidth}
            \centering
            \includegraphics[width=0.9\textwidth]{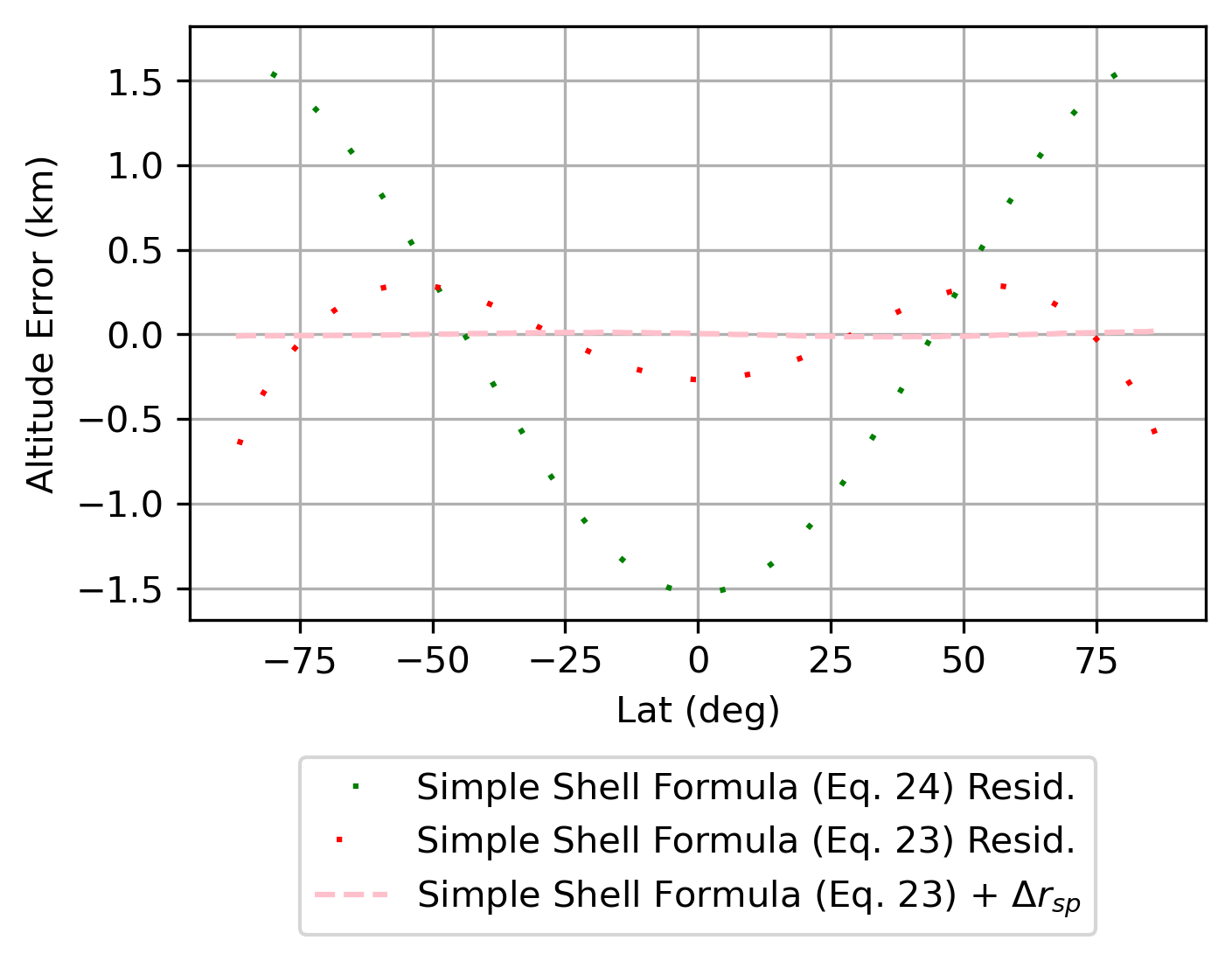}
              \caption{Fit Residuals}
        \end{subfigure}
        \caption{Genetic Algorithm  Geometry Comparison Example 2 ($\mathbf{i = 87 [deg], N_o = 5, N_{so} = 5, N_c = 2}$)}
    \label{fig:inc87fitga}
    \end{figure}

    To rule out a difference in mean element generation methodology, we then used a genetic algorithm allowing each run to adjust values for mean a, e, and $\omega$ (if the equation included it) to minimize root mean squared error between the interpolated shell mean and each of the shell geometry equations. Inclination was assumed to remain fixed at the mean values calculated using DSST. As is evident in Figures \ref{fig:inc40fitga} and \ref{fig:inc87fitga}, even allowing for arbitrary input parameters, Equations \ref{range} and \ref{radius_lat} have remaining residuals of hundreds of meter to just over a kilometer while Equation \ref{radius_lat_drsp} is able to almost perfectly match the reference curve.
    
\begin{figure}[htbp]
        \centering
        \begin{subfigure}[t]{.45\textwidth}
            \centering
            \includegraphics[width=0.9\textwidth]{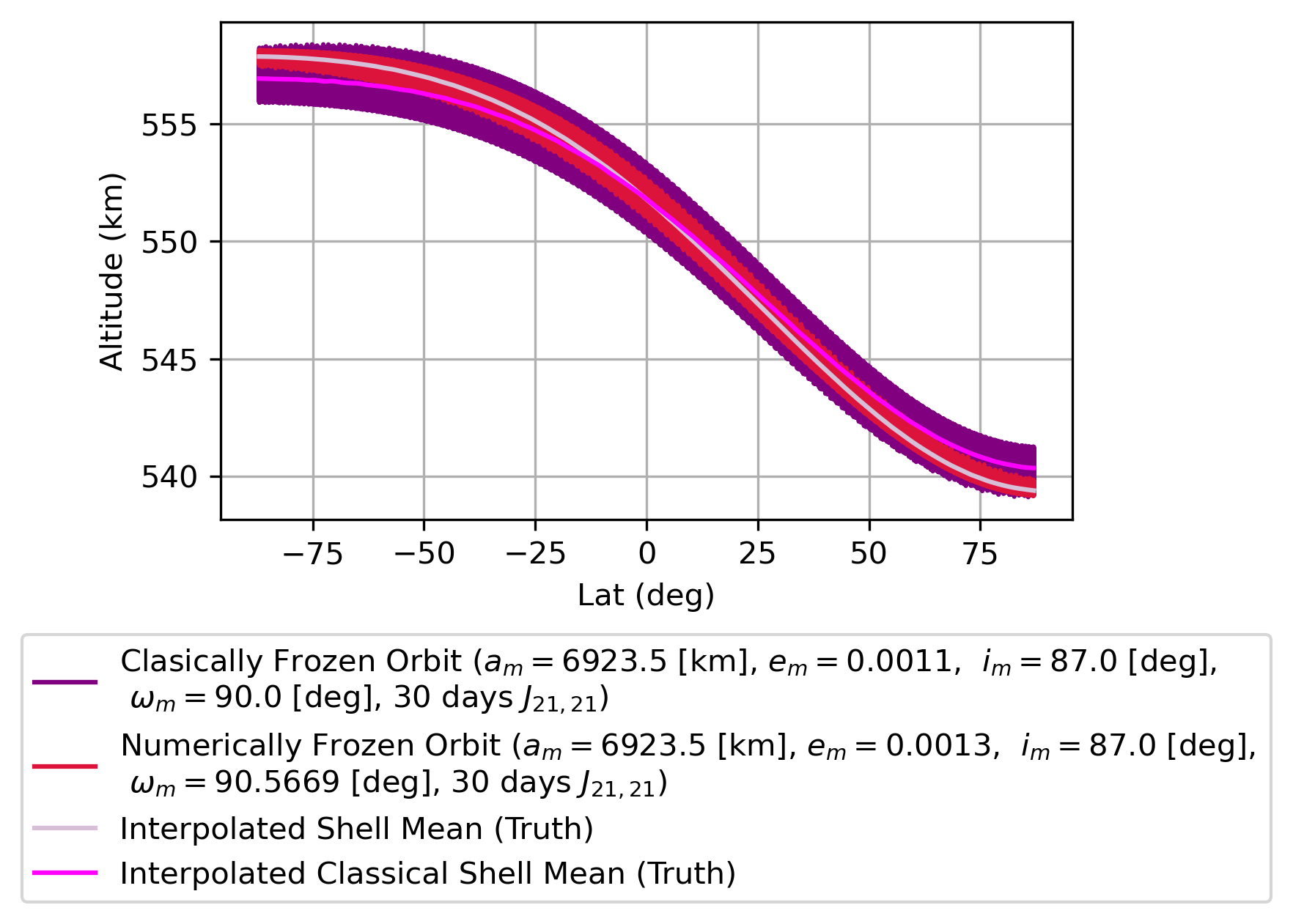}
            \caption{Equation Fit}
        \end{subfigure}
        \begin{subfigure}[t]{.45\textwidth}
            \centering
            \includegraphics[width=0.9\textwidth]{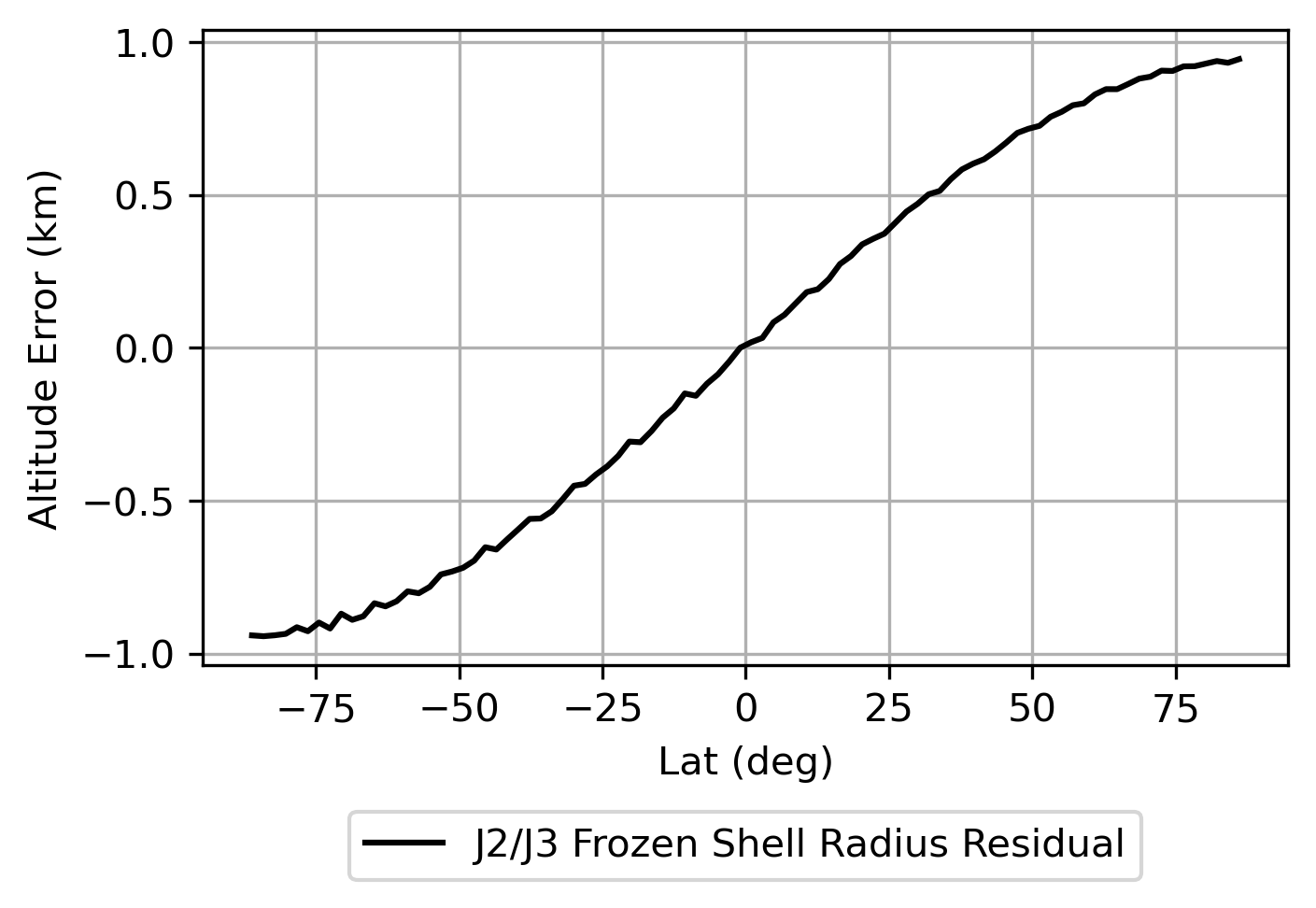}
              \caption{Fit Residuals}
        \end{subfigure}
        \caption{Classical vs. Numerical Orbit Freezing Geometry Comparison Example 2 ($\mathbf{i = 87 [deg], N_o = 5, N_{so} = 5, N_c = 2}$)}
    \label{fig:J2J3Comp87}
    \end{figure}

    We next explored the impact on shell center-line geometry and residuals from using $J_2/J_3$ classical frozen orbit eccentricity and argument of perigee values versus those calculated numerically. Note that in this example, both options are evaluated for a single seed trajectory, rather than for a full constellation. Errors for center-line curves were small for the low inclination shells (tens of meters), but grew to nearly $\pm 1$ km for the $87$ degree shell in Figure \ref{fig:J2J3Comp87}.
    
    \begin{figure}[htbp]
        \centering
        \begin{subfigure}[t]{.45\textwidth}
            \centering
            \includegraphics[width=0.9\textwidth]{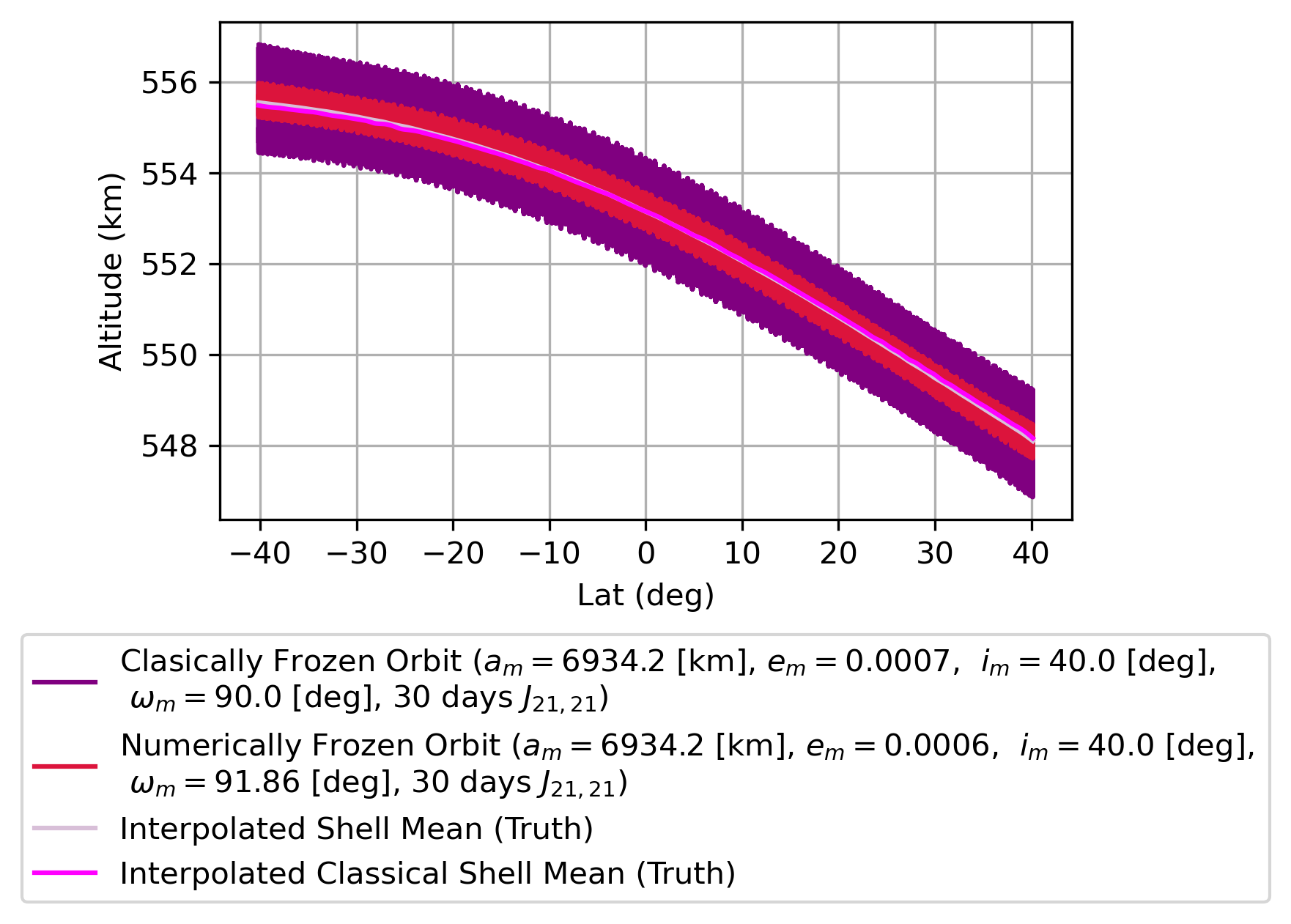}
            \caption{Equation Fit}
        \end{subfigure}
        \begin{subfigure}[t]{.45\textwidth}
            \centering
            \includegraphics[width=0.9\textwidth]{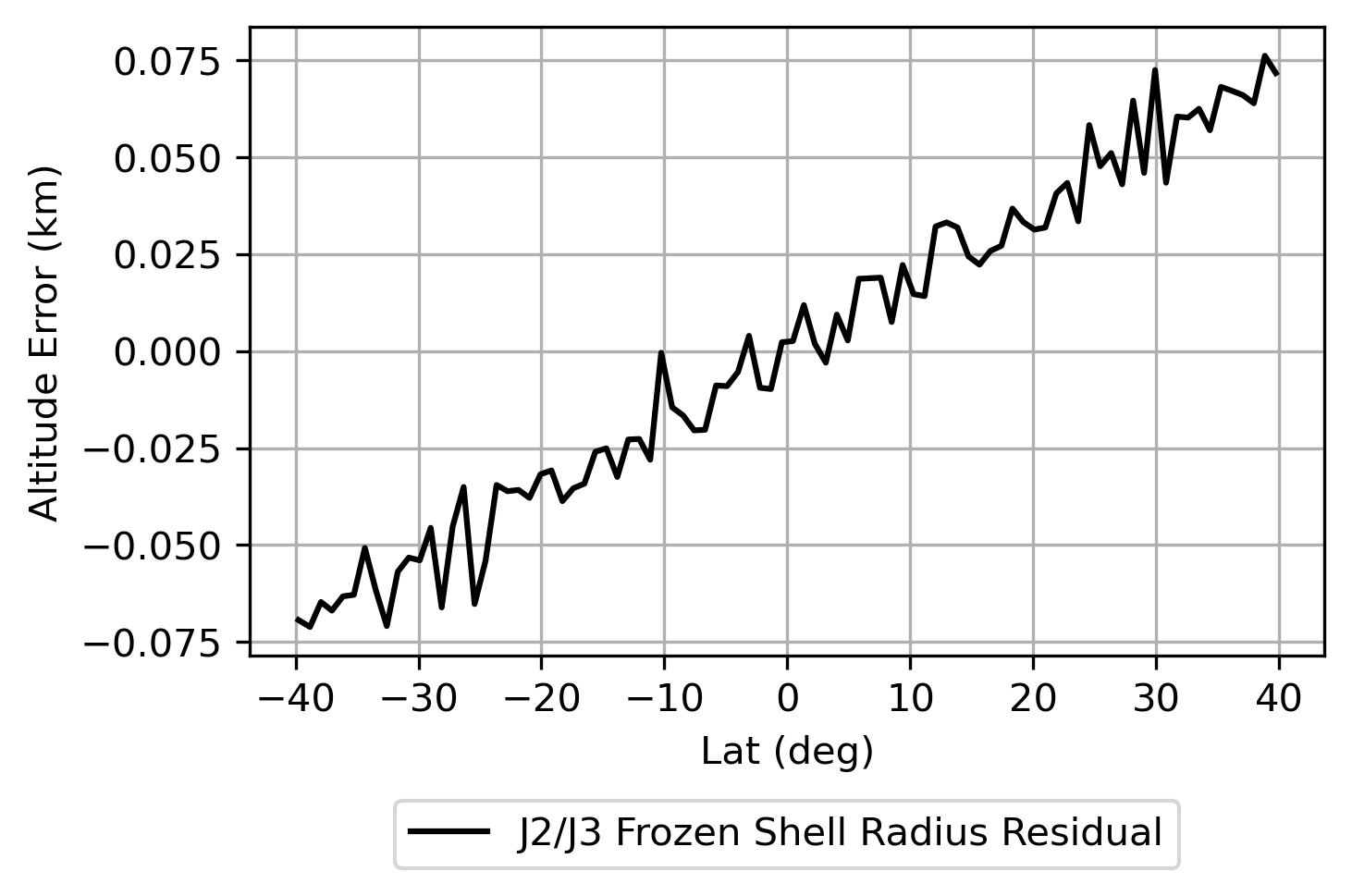}
              \caption{Fit Residuals}
        \end{subfigure}
        \caption{Classical vs. Numerical Orbit Freezing Geometry Comparison Example 1 ($\mathbf{i = 40 [deg], N_o = 5, N_{so} = 5, N_c = 2}$)}
    \label{fig:J2J3Comp40}
    \end{figure}  
    
    Based on these results we conclude that for the purpose of shell geometry estimation, it is necessary to include the short term periodic effects due to $J_2$ and that doing so is able to reproduce shell shape and altitude to an accuracy of hundreds of meters across several examples. When estimating shell shapes for stacking purposes, using frozen eccentricities incorporating higher degree zonal effects improves accuracy as compared to $J_2/J_3$ frozen eccentricities. Nevertheless, the difference may be small enough to reasonably ignored for estimating shell center-line geometry for some purposes.  However, incorporation of higher fidelity orbit freezing is very necessary for the stability of the resulting orbits, as is visible in Figures \ref{fig:J2J3Comp40} and \ref{fig:J2J3Comp87}.

    \subsection{Shell Width Estimation}
    
    To investigate the effect of the time varying orbital elements on the osculating $r(\phi)$ we compute a first-order linearization of the relationship in Eq.~\ref{radius_lat_drsp}. The linearized error propagation relationships are as follows:
    \begin{equation}
    \sigma_r^2 =\left|\frac{dr}{da}\right|^2\sigma_a^2+\left|\frac{dr}{de}\right|^2\sigma_e^2+\left|\frac{dr}{d\omega}\right|^2\sigma_\omega^2+\left|\frac{dr}{d i}\right|^2\sigma_i^2
    \end{equation}
    where the $\sigma$ terms represent the error in the mean orbital elements and osculating radius parameters. Note that this expression assumes Gaussian errors in the mean orbital elements and translates these errors into Gaussian errors in osculating $r$. Therefore the interpretation is that if we take the shell width to be given by $3 \sigma_r$ then it should include 99.7 \% of all the errors due to $J_2$ variations. However, our full dynamic model includes higher-order terms over long propagation intervals that are neglected by Eq.~\ref{radius_lat_drsp}. We have found that $6 \sigma_r$ can reasonably account for these unmodeled effects (zonal terms higher than $J_2$ and sectoral and tesserial gravity terms) over a 30 day propagation period. Such a model can be used to account for estimation and control errors and for a case that doesn't freeze the satellite orbit but rather controls the trajectory to achieve the desired shell width. This work uses this relationship to estimate the shell width at given altitudes.

    The full expressions used for computing the shell with is shown in the appendix but we will examine the behavior of the radius as a function of latitude without the short period terms to gain an intuitive understanding. Therefore, assuming $\omega=\pi/2$ and that $\Delta r_{sp}=0$ (note this just for the expression below. We use the full expression in our shell width estimation numerical results), we have the following:
        $$\frac{dr}{da}=-\frac{\sin\left(\mathrm{i}\right)\,\left(e^2-1\right)}{\sin\left(\mathrm{i}\right)+e\,\sin\left(\phi \right)}= \begin{cases}
          1-e^2, & \text{with}\ \phi=0 \\
          1-e, & \text{with}\ \phi=i
        \end{cases}$$

    $$\frac{dr}{de}=-\frac{a\,\sin\left(\mathrm{i}\right)\,\left(\sin\left(\phi \right)\,e^2+2\,\sin\left(\mathrm{i}\right)\,e+\sin\left(\phi \right)\right)}{{\left(\sin\left(\mathrm{i}\right)+e\,\sin\left(\phi \right)\right)}^2}= \begin{cases}
          -2\,a\,e, & \text{with}\ \phi=0 \\
          -a, & \text{with}\ \phi=i
        \end{cases}$$
    
    $$\frac{dr}{d\omega}=-\frac{a\,e\,{\sin\left(\mathrm{i}\right)}^2\,\left(e^2-1\right)\,\sqrt{\frac{{\sin\left(\mathrm{i}\right)}^2-{\sin\left(\phi \right)}^2}{{\sin\left(\mathrm{i}\right)}^2}}}{{\left(\sin\left(\mathrm{i}\right)+e\,\sin\left(\phi \right)\right)}^2}= \begin{cases}
          a\,e\,\left(1-e^2\right), & \text{with}\ \phi=0 \\
          0, & \text{with}\ \phi=i
        \end{cases}$$
    
    $$\frac{dr}{d i}=-\frac{a\,e\,\cos\left(\mathrm{i}\right)\,\sin\left(\phi \right)\,\left(e^2-1\right)}{{\left(\sin\left(\mathrm{i}\right)+e\,\sin\left(\phi \right)\right)}^2}= \begin{cases}
          0, & \text{with}\ \phi=0 \\
          ae\cot\left(i\right)\frac{e-1}{1+e}, & \text{with}\ \phi=i
        \end{cases}$$
    Note that if we let $\phi=0$, the derivative term for $e$ is $-2a e$ and if we let $\phi=i$ the derivative term for $e$ is $-a$.
    The effect of variation in the orbital elements can be taken into account using these first order relationship for small deviations. To consider the effect of $e$ on the slope of the radius function we take the derivative of $r$ with respect to $\phi$ by substituting $\theta=\mathrm{arcsin}(\sin(\phi)/\sin(i))$. We compute the following
        \begin{equation}
        \frac{d r}{d \phi}=\frac{a\,e\,\sin\left(w-\mathrm{arcsin}\left(\frac{\sin\left(\phi \right)}{\sin\left({i}\right)}\right)\right)\,\cos\left(\phi \right)\,\left(e^2-1\right)}{\sin\left({i}\right)\,\sqrt{1-\frac{{\sin\left(\phi \right)}^2}{{\sin\left({i}\right)}^2}}\,{\left(e\,\cos\left(w-\mathrm{arcsin}\left(\frac{\sin\left(\phi \right)}{\sin\left({i}\right)}\right)\right)+1\right)}^2}
    \end{equation}
    or for $\omega=\pi/2$ we have the following simplification,
    $$\frac{d r}{d \phi}=\frac{a\,e\,\cos\left(\phi \right)\,\sin\left({i}\right)\,\left(e^2-1\right)}{{\left(\sin\left({i}\right)+e\,\sin\left(\phi \right)\right)}^2}
    $$
    Note that if $e=0$ then the slope of this curve is zero as expected and the slope scales with eccentricity and semi-major axis. 
    
\begin{figure}[htbp]
        \centering
        \begin{subfigure}[t]{.45\textwidth}
            \centering
            \includegraphics[width=\textwidth]{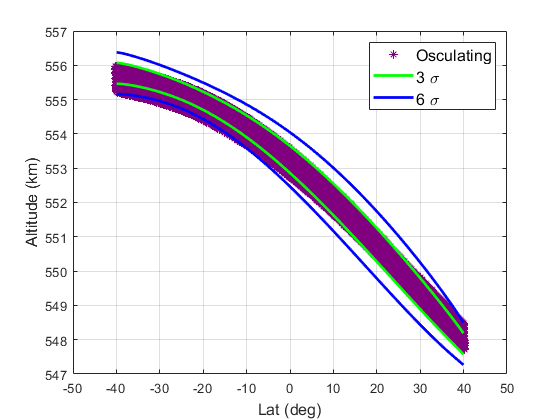}
            \caption{$\mathbf{i = 40 [deg]}$}
        \end{subfigure}
        \begin{subfigure}[t]{.45\textwidth}
            \centering
            \includegraphics[width=\textwidth]{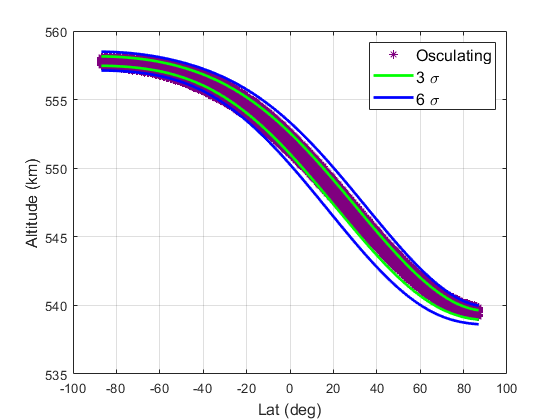}
              \caption{$\mathbf{i = 87 [deg]}$}
        \end{subfigure}
        \caption{Estimated vs. Actual Shell Widths for Sample Satellites}
    \label{fig:shellwidth}
    \end{figure}

    Figure \ref{fig:shellwidth} shows results of this width estimation approach for the seed satellite trajectory for the 40 degree and 87 degree shells used in the previous modeling examples, propagated for 30 days under a 21 by 21 gravity field model. Mean elements were taken from DSST mean elements for the propagated trajectory for the same satellite.
    
    \subsection{Visualizing Shells in the Latitude-Altitude Plane and Key Implications for Layer Stacking}
    
\begin{figure}[htb]
        \centering
        \includegraphics[width=0.65\textwidth]{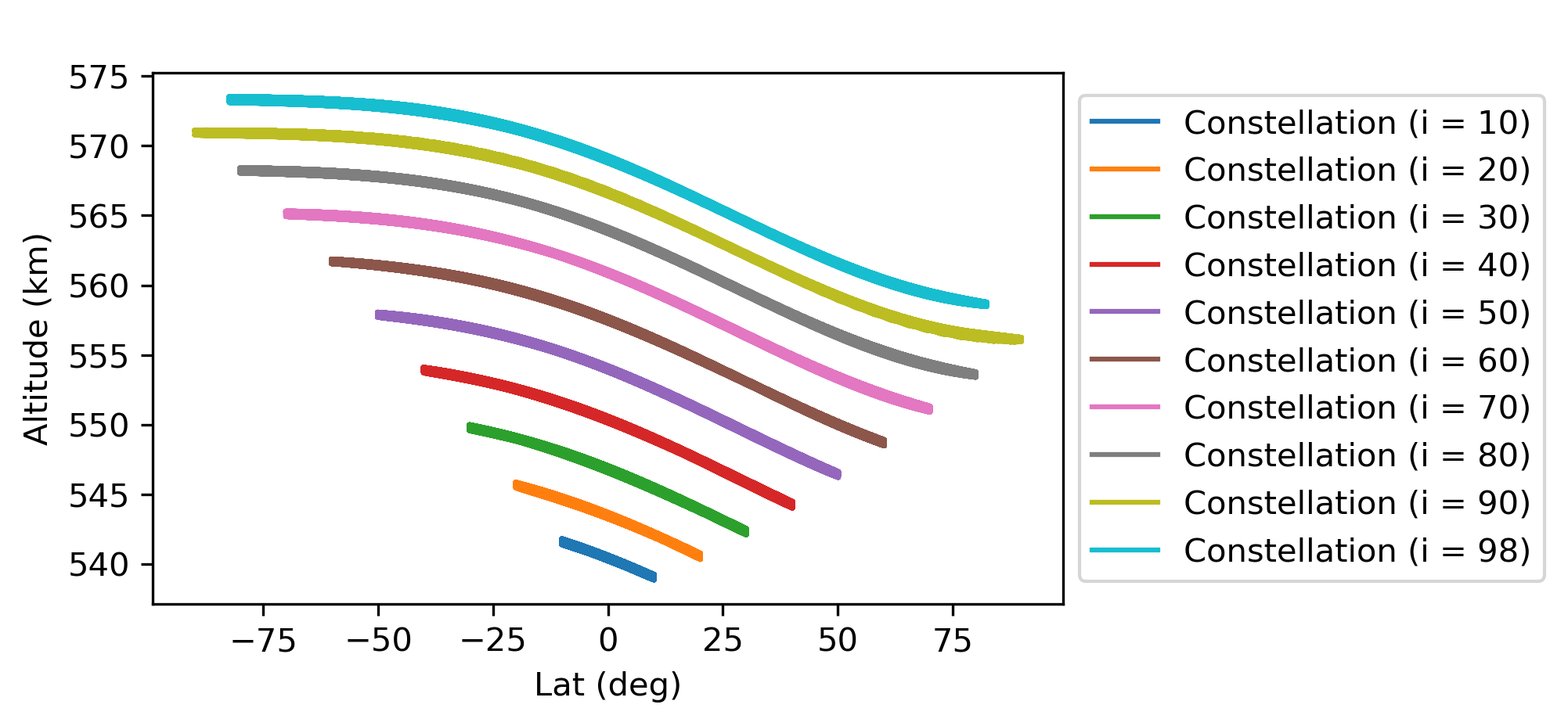}
        \caption{$J_2$-$J_3$ Frozen shells propagated under a $J_{3,0}$ model for 30 days. The shells show inclination-dependent shape and tessellate well}
        \label{fig:J2J3Shells}
    \end{figure}

\begin{figure}[htb]
        \begin{subfigure}[t]{.45\textwidth}
            \centering
            \includegraphics[width=0.95\textwidth]{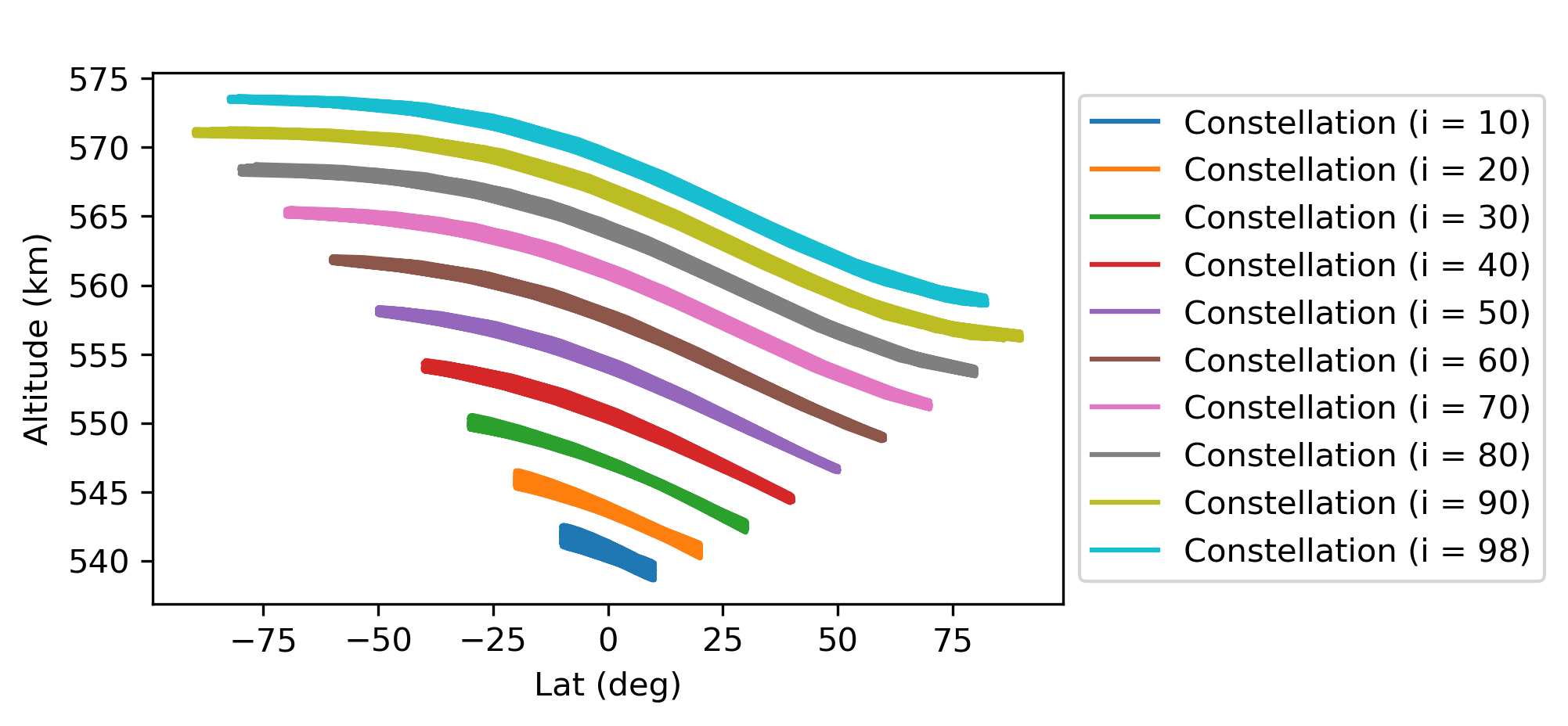}
            \caption{1 Day}
        \end{subfigure}
        \begin{subfigure}[t]{.45\textwidth}
            \centering
            \includegraphics[width=0.95\textwidth]{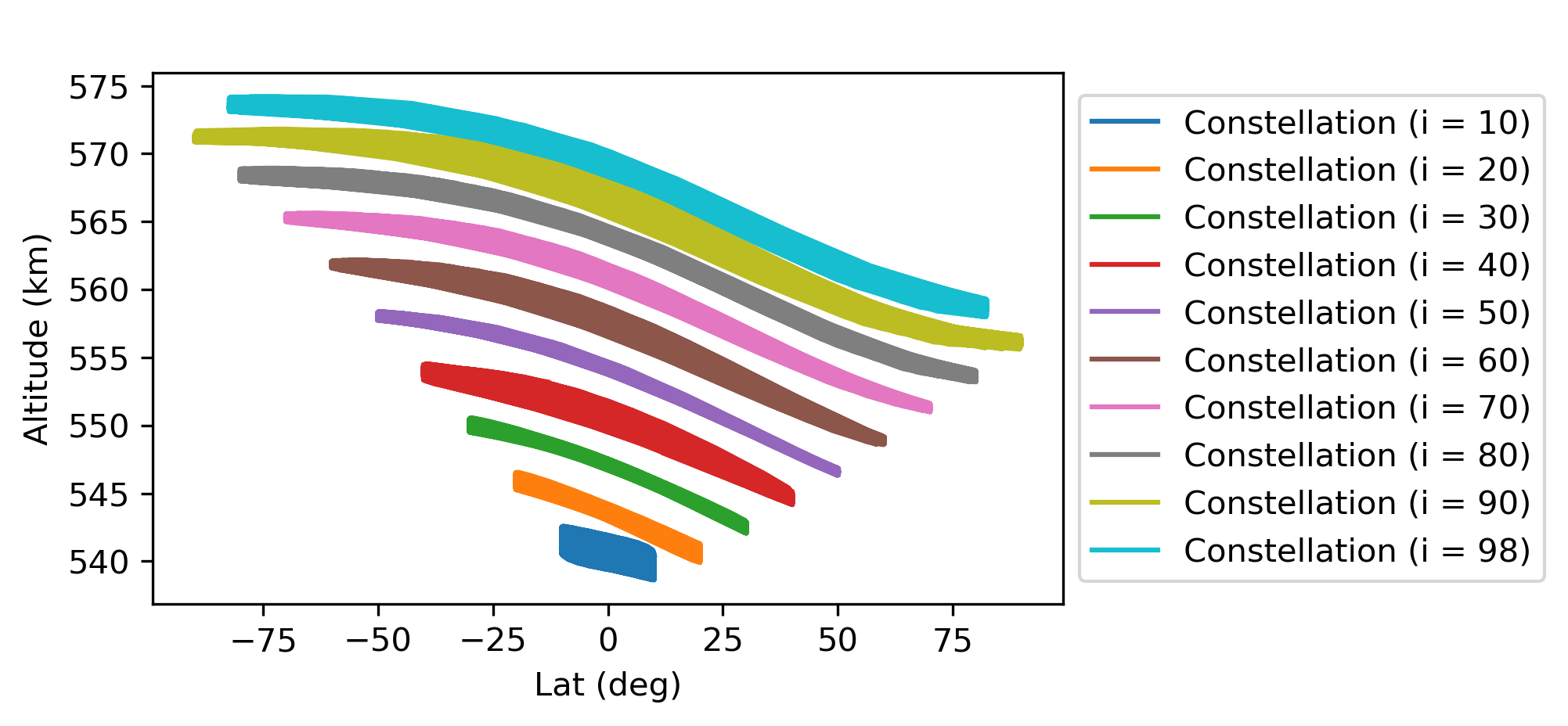}
            \caption{7 Days}
        \end{subfigure}
        \\
        \begin{subfigure}[t]{.45\textwidth}
            \centering
            \includegraphics[width=0.95\textwidth]{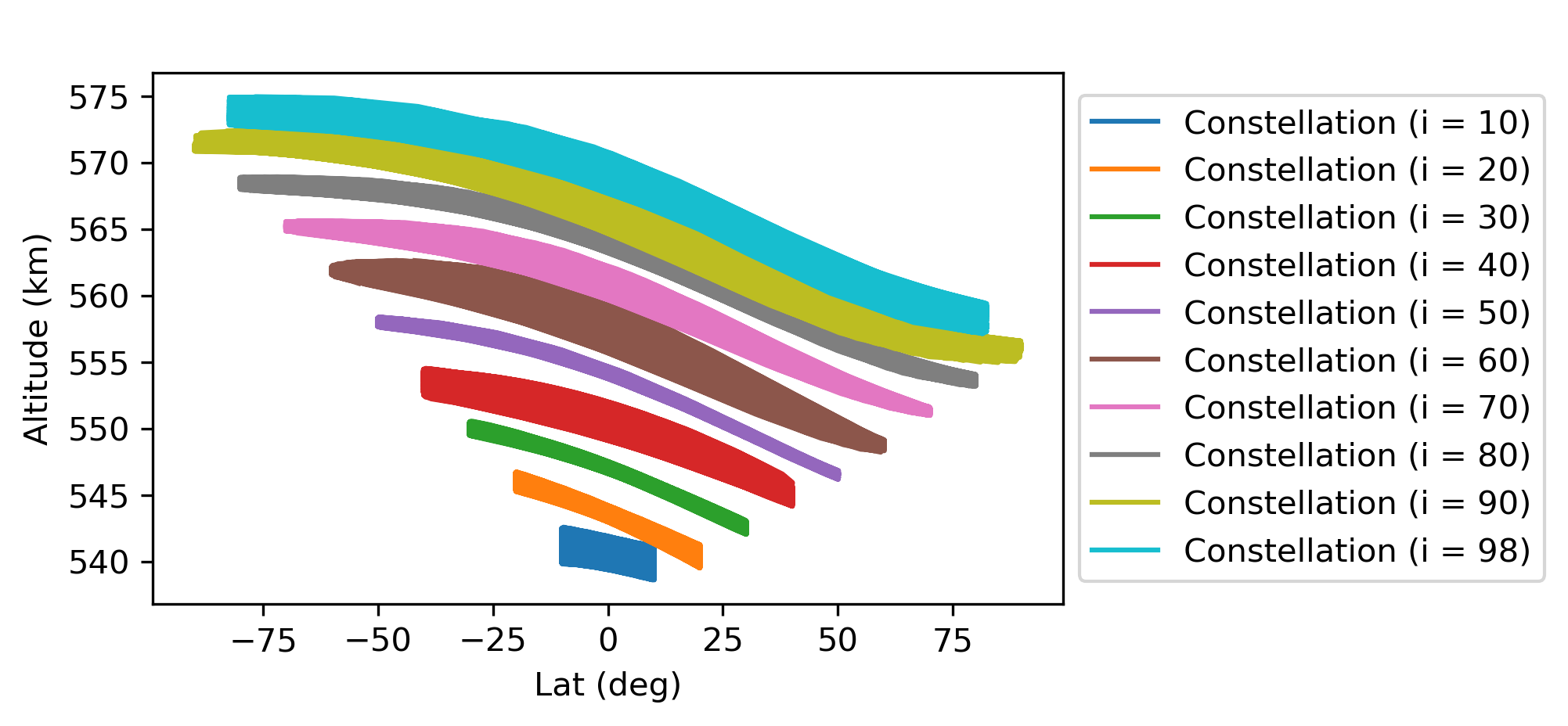}
            \caption{14 Days}
        \end{subfigure}
        \begin{subfigure}[t]{.45\textwidth}
            \centering
            \includegraphics[width=0.95\textwidth]{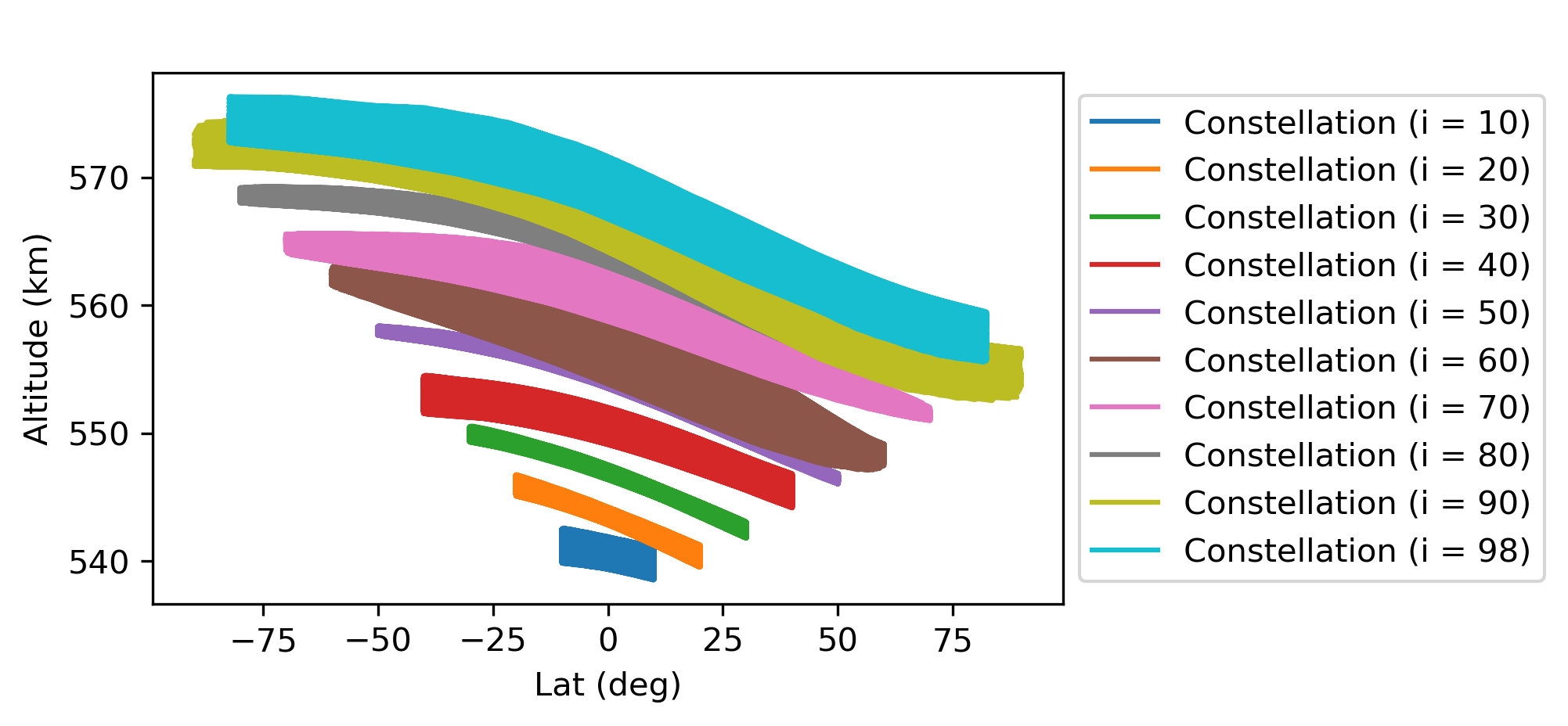}
            \caption{30 Days}
        \end{subfigure}
        \caption{$J_2$-$J_3$ Frozen shells propagated under a $J_{21,21}$ model. The shells lose coherence over time.}
        \label{fig:J2J3ShellsFullForce}
    \end{figure}

    Results generated using either of the presented 2D-LFC definition methods or a simplified shell model can be visualized in the latitude-altitude plane. Separation between shells in this space provides intuitive assurance that over the time period propagated, the shells with not intersect and do not pose an inter-shell conjunction hazard. The shape of shells is primarily a function of inclination and eccentricity. As seen in Figure \ref{fig:J2J3Shells} and demonstrated mathematically in Reference~\cite{Bombardelli2021}, shells generated to be classically frozen under a $J_{3,0}$ model are arbitrarily stable and thin, subject only to numerical error. However, as seen in Figure \ref{fig:J2J3ShellsFullForce}, $J_2$-$J_3$ shells are not sufficiently stable when propagated under a more realistic geopotential and rapidly lose coherency. While shells can be defined to account for higher order and/or degree geopotentials, as described previously, this leads to additional shell thickness that increases with time and magnifies differences in shell shape as a result of inclination. 
    
\begin{figure}[htbp]
        \centering
        \includegraphics[width=0.45\textwidth]{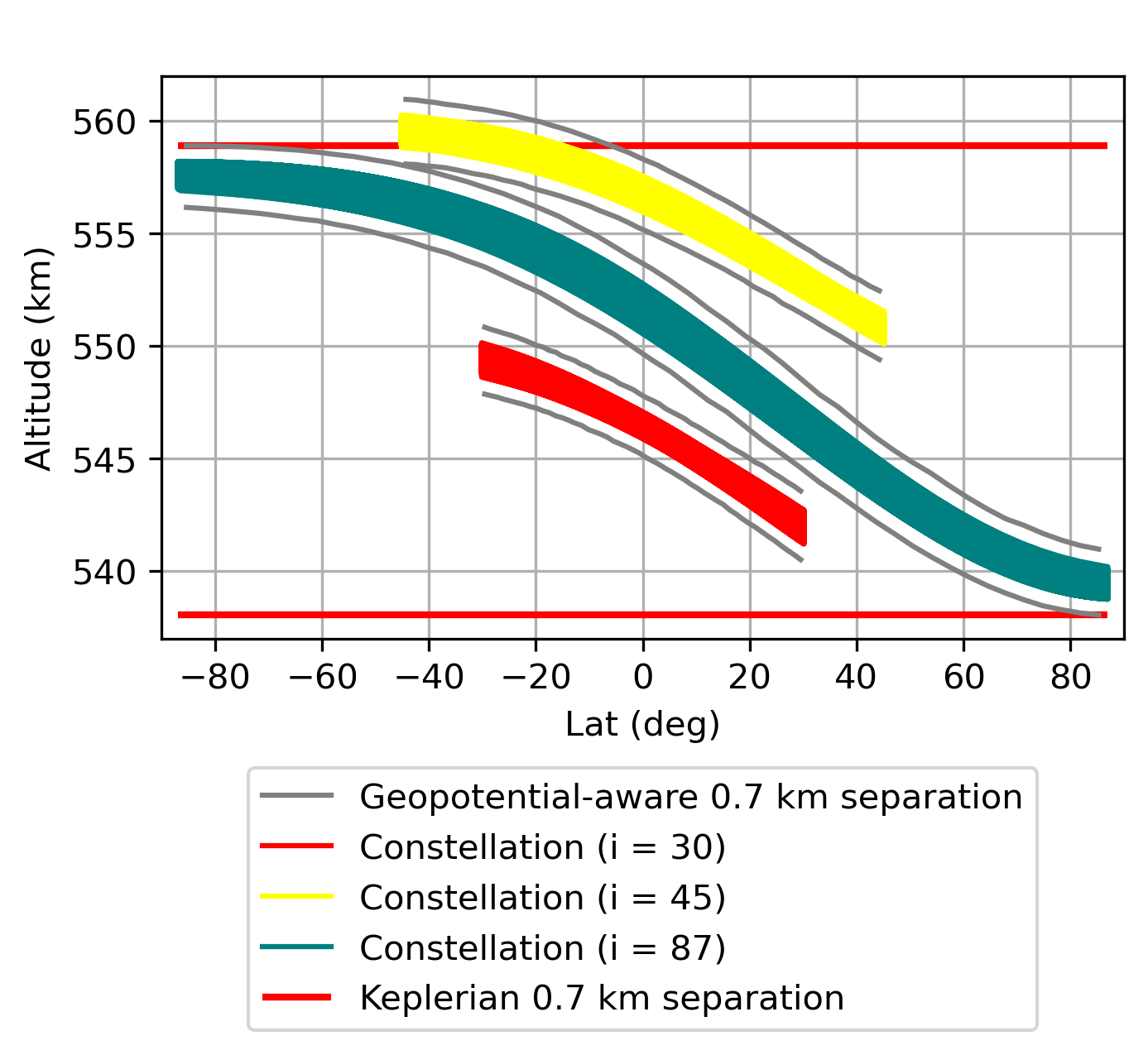}
        \caption{Shells generated using frozen orbits can nest closer than the osculating variation in semi-major axis for each shell and still maintain coherency and safety (zonal freezing method, 30 day propagation, \begin{math}J_{21,21}\end{math} model).}
        \label{fig:shellnesting}
    \end{figure}

    Shells designed using either of the above methods are stable over longer period of times, and preserve similar separation geometry. As seen in Figure \ref{fig:shellnesting}, frozen coordinated shells can under many circumstances be placed closer than an arbitrary Keplerian minimum separation distance between shells while preserving the same actual minimum separation distance between shells! This is a very important result for ensuring simultaneous orbital efficiency and safety.

\begin{figure}[htb]
        \begin{subfigure}[t]{.45\textwidth}
            \centering
            \includegraphics[width=0.95\textwidth]{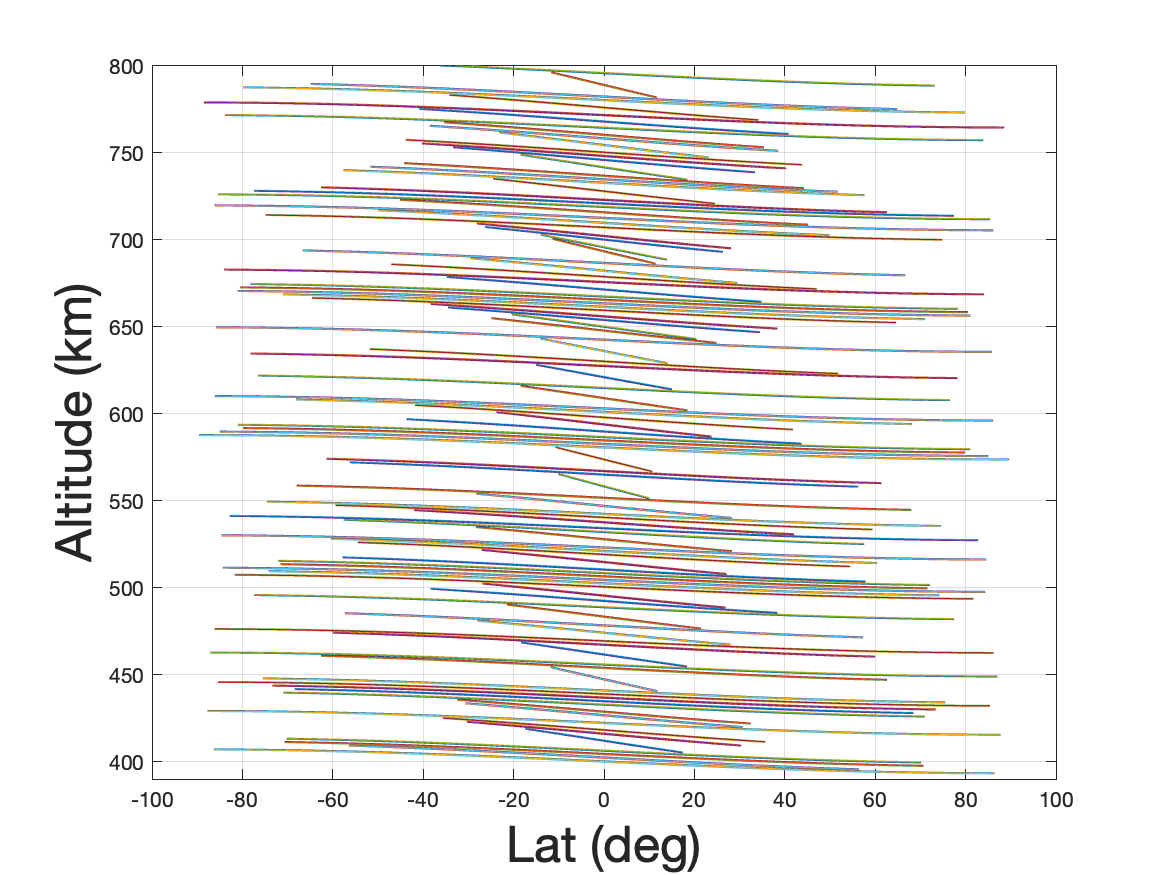}
            \caption{Randomly Ordered Inclinations}
        \end{subfigure}
        \begin{subfigure}[t]{.45\textwidth}
            \centering
            \includegraphics[width=0.95\textwidth]{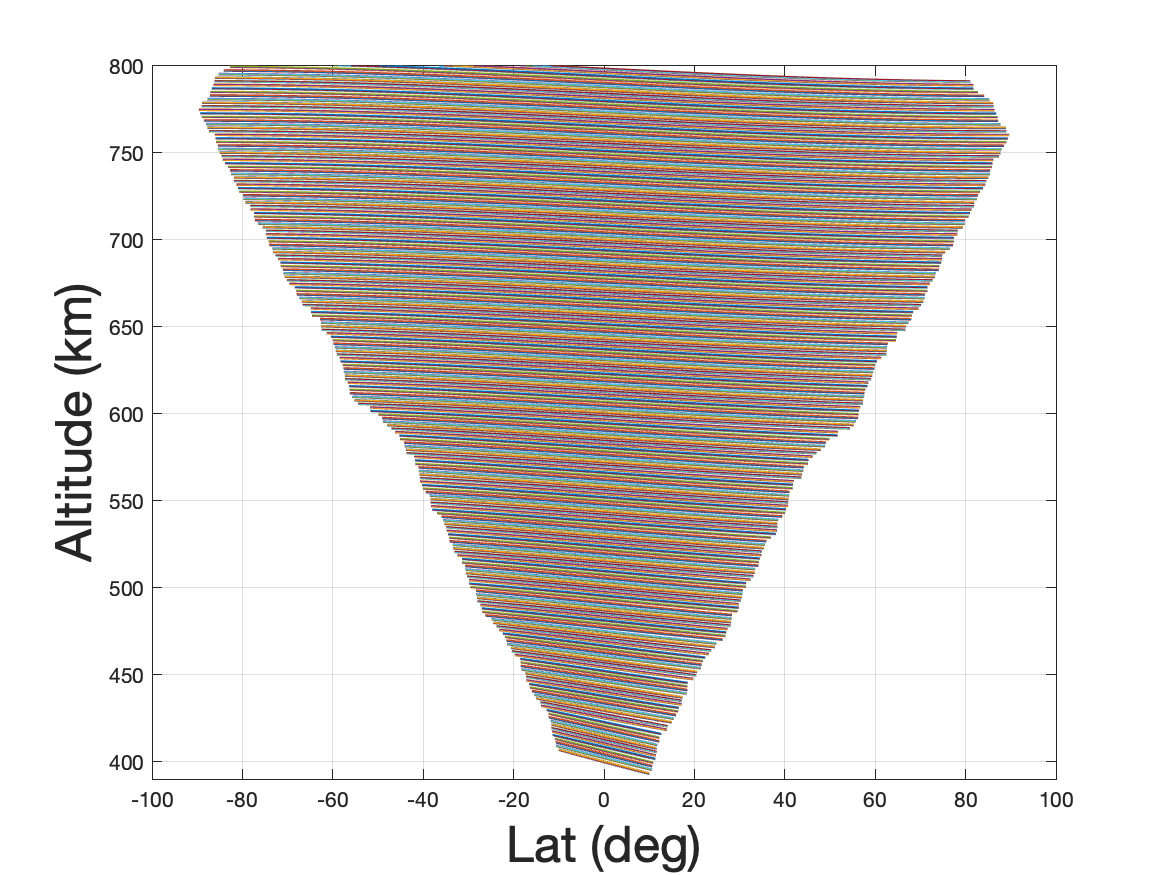}
            \caption{Sequentially Ordered Inclinations}
        \end{subfigure}
    \caption{A Comparison of Random and Sequential Inclination Shell Layering Approaches using Equation \ref{radius_lat_drsp}}
    \label{fig:stack}
    \end{figure}

    Use of coordinated sequential inclinations can result in significant increases to the number of admissible shells when compared to an alternative with randomly chosen inclinations spaced to ensure non-overlap between shells. In Figure \ref{fig:stack}, two notional configurations for the equatorial altitude range from 400-800 km are considered to demonstrate this effect, each preserving separation between shell center-lines. Shell center-lines are estimated using Equation \ref{radius_lat_drsp} rather than explicitly calculated and propagated. In the left figure, each shell has a randomly selected inclination (assumed to be in the interval [10, 100] degrees), while in the right figure, shells are placed in strictly ascending order of inclination. Ten satellites were modeled for each shell, with each satellite having random per-satellite mean eccentricity values drawn from a uniform random distribution between 0.00102 and 0.0012, a random inclination offset from the shell inclination of between 0.1 and 1 degrees (drawn from a uniform random distribution), and an argument of perigee of 90 degrees with per satellite error of between 2 and 20 degrees (again sampled from a uniform random distribution). The sequential strategy is able to fit 103 more shells (196 vs. 93) when compared to the random approach. Assuming every shell uses a frozen eccentricity rather than a random small eccentricity would reduce but not eliminate the penalty associated with random shell order. This ordering and spacing is not presented as a proposal that we believe should be specifically implemented in LEO, but rather than to be indicative of gains to capacity from coordination of shell inclinations.
    
\section{Results}\label{Results}

    Several implications of these stacking techniques are explored in this section. In this initial smaller examples, 2D-LFC with parameters $N_o = 5, \ N_{so} = 5, \ N_c = 2$ are used with various mean inclinations and altitudes. When satellites or slots are said to be propagated under a full-force model, this includes a 21x21 geopotential using the EIGEN-6S gravity field model, solar radiation pressure (isotropic radiation, $c_r = 1, \ a_{srp} = 28$ m), atmospheric drag ($c_d = 2.2$, $a_{drag} = 15$ m, DTM2000 model using CSSI space weather data), solar and lunar Earth solid tides, and lunar and solar third body attraction (based on DE-440 ephemerides). All starting epochs are assumed to be 1 January 2022 0:00:00.000 UTC. Two examples are then demonstrated, drawing on proposed constellations.  These examples are chosen somewhat arbitrarily from examples of large constellations proposed at nearby altitudes. No statement is intended or should be inferred from the selection of a specific constellation pair for inclusion in this section.
    
    Shell widths and the separation distance necessary between shells depends on multiple factors including the chosen 2D-LFC used to generate the shell, satellite locations relative to the nominal center of each slot, orbital state estimation and control accuracy, desired additional safety margin, satellite physical properties, and the frequency of orbital station-keeping desired for the constellation. 
    
    \subsection{Implications of Inclination Ordering on Orbital Separation}

\begin{figure}[htbp]
        \centering
        \begin{subfigure}[t]{.45\textwidth}
            \centering
            \includegraphics[width=0.9\textwidth]{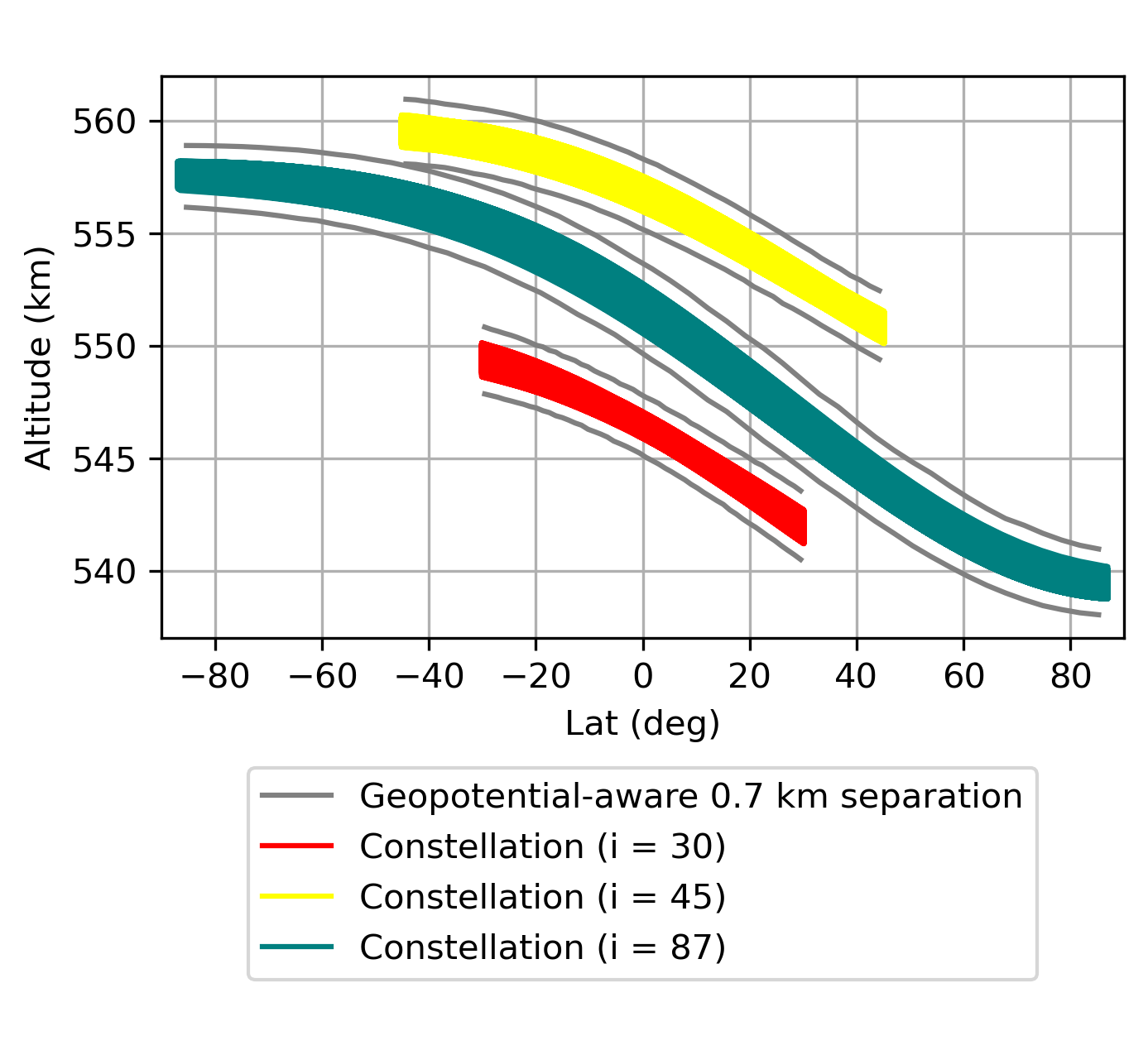}
            \caption{Non-sequential inclination shells (same example as Figure \ref{fig:shellnesting})}
            \label{fig:NonSequentialShells}
        \end{subfigure}
        \begin{subfigure}[t]{.45\textwidth}
            \centering
            \includegraphics[width=0.9\textwidth]{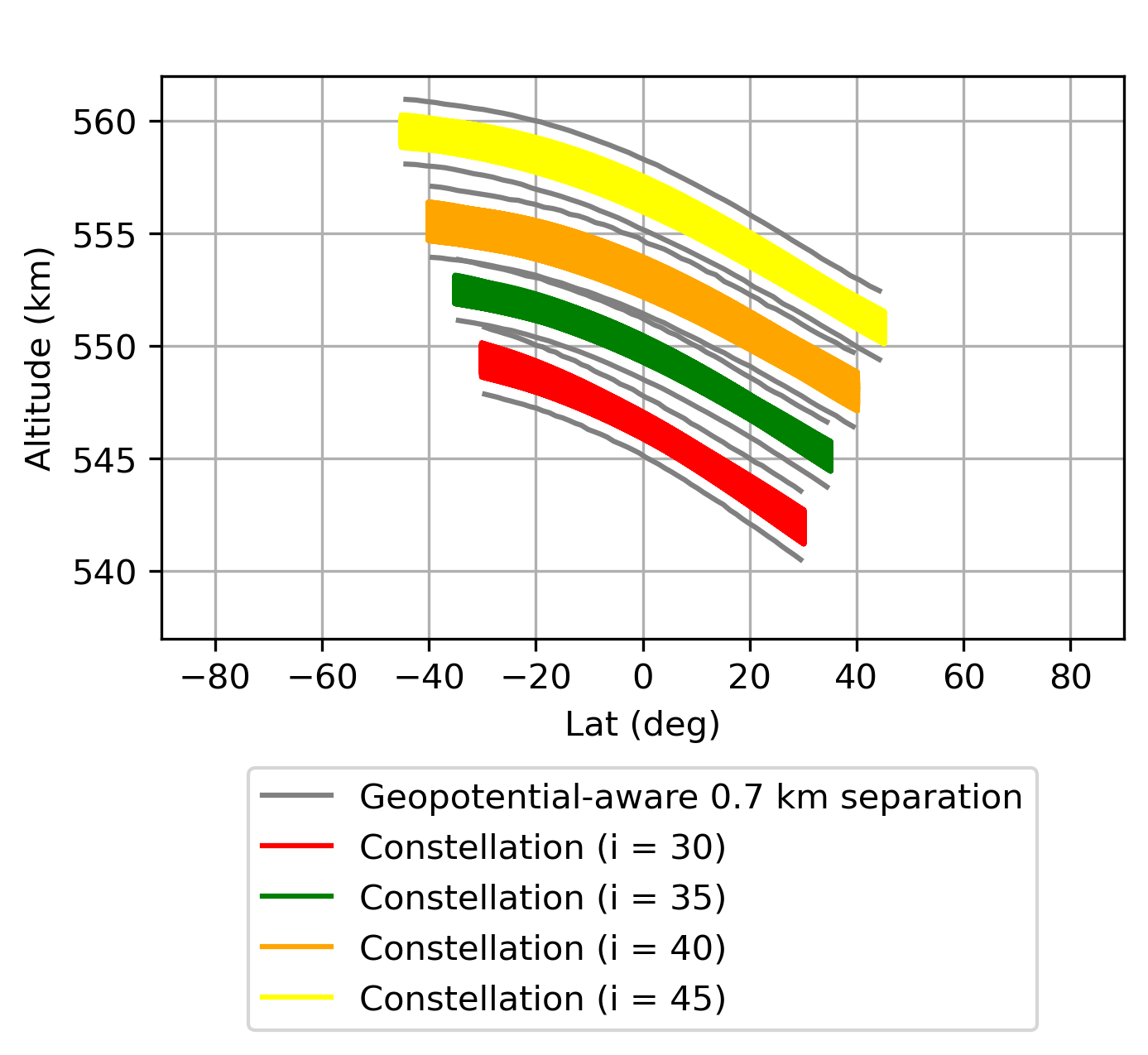}
            \caption{Sequential inclination shells}
            \label{fig:SequentialShells}
        \end{subfigure}
        \caption{Shells with sequential inclinations tessellate better than with very different inclinations (zonal freezing method, 30 day propagation, \begin{math}J_{21,21}\end{math} model).}
    \label{fig:shellstacking}
    \end{figure}

    Shell shape in the latitude-altitude plane under a geopotential-only model is largely a function of eccentricity and inclination. Stacking shells with similar inclinations in an ascending or descending sequence inclinations (or at least similar inclinations) allows for dense stacking that exploits the inherent tessellating effect. As see in Figure \ref{fig:shellstacking}, two sequential shells can fit within the space occupied by a single shell with a very different inclination. This ordering is an important results and represents a trade-off for constellation operators where coordination could increase capacity, particularly in LEO regions characterized by significant demand for both Sun-synchronous and lower-inclination orbits.
    
    \subsection{Minimum Distance Comparison Between Keplerian and Osculating 2D-LFC}
    
    To verify that the generated slots within a shell preserve collision avoidance, a constellation was chosen from a database of high capacity 2D-LFCs and checked for conjunction events over a 30 day period, after being numerically propagated using the same $J_{21,0}$ geopotential used to define the adjusted 2D-LFC using the method in Section \ref{geomethod}. The chosen constellation was placed at 600 km in altitude at an inclination of 60 degrees and has 2D-LFC parameters $N_o = 19, \ N_{so} = 26, \ N_c = 6$. The elements $\Omega$, and $M$ were set to 0 for the seed orbit. As calculated using the method from \cite{Avendano2021}, this constellation has a Keplerian minimum separation distance of 1.408 degrees. This corresponds to a separation of approximately 171.4 km (for this separation and altitude the difference between arc length and cord length is minimal). A comparison was conducted between every pair of satellites every five seconds, based on the premise that even if two satellites moved directly towards one another for a five second period at their orbital velocities, they would be unable to collide before the next screening time. The closest approach identified between slot centers is 130.556 km. This distance decreases by 352 meters over the course of the simulation, or about 0.27\% of the initial separation distance.
 
    \subsection{Example: Zonal-Only Approach, Low Altitude}
    
       We now demonstrate the zonal method for a set of nearby orbital shells. Both Amazon and SpaceX have proposed expansions to their very large LEO constellations. As part of SpaceX's Gen2 Configuration 1 Proposal\footnote{SpaceX's Gen2 proposal can be viewed at \url{https://licensing.fcc.gov/cgi-bin/ws.exe/prod/ib/forms/reports/swr031b.hts?q_set=V_SITE_ANTENNA_FREQ.file_numberC/File+Number/\%3D/SATAMD2021081800105\&prepare=\&column=V_SITE_ANTENNA_FREQ.file_numberC/File+Number}.}, it requested permissions to operate shells at 604 and 614 km.  Amazon has received approval for a shell at 610 km that would be expanded to have twice as many satellites as part of its most-recent FCC application\footnote{Amazon's proposal can be viewed at \url{http://licensing.fcc.gov/cgi-bin/ws.exe/prod/ib/forms/reports/swr031b.hts?q_set=V_SITE_ANTENNA_FREQ.file_numberC/File+Number/\%3D/SATLOA2021110400145\&prepare=\&column=V_SITE_ANTENNA_FREQ.file_numberC/File+Number}.}. Overlaps between these shells are particularly concerning because Amazon's shells are prograde while these two SpaceX shells are retrograde, meaning any collisions would likely occur at an especially high relative velocity in a very dense region of LEO. In this subsection, we demonstrate how the zonal approach described in this paper could be used to maintain minimum separation between these shells.  This example has not been discussed with either operator, and no claim is made as to whether implementing this method would be feasible given the concept of operations and spacecraft capabilities of either company.
       
       This example is a bit unusual because of the retrograde orbits proposed by SpaceX, one of which lies very near critical inclination with consequences for frozen orbit stability. As discussed in Section \ref{ssec:simpleshellmodel}, the direction of the slope of shells in the latitude-altitude space is dependent on the argument of perigee about which the curve is frozen.  In this case, the GA prioritizes a frozen orbit for the SpaceX 614 km shell that slopes the opposite direction of the other two shells (associated with the frozen orbit nearer to $\omega = 3\pi/2$, for which the seed orbit has an initial osculating argument of perigee of 4.34 radians).  This shell could be raised to a higher nominal altitude, although it would then overlap a proposed Amazon Kuiper shell at 630 km, or the GA can be constrained to use a solution nearer to $\omega = \frac{\pi}{2}$, for which the seed orbit has an initial osculating argument of perigee of 2.042 radians).  In this case, the new solution is less stable, but is compatible with the other two shells. The shell is initialized about 5 km higher than the reference altitude to leave ample separation from both these shells and the proposed Amazon shell at 630 km.

        \begin{table*}[htbp]
        \resizebox{\textwidth}{!}{%
        \begin{tabular}{lllllll}
        \textbf{Name}                         & \textbf{Nom. Altitude (km)} & \textbf{Inclination (deg)} & $\mathbf{N_{sat}}$ & $\mathbf{N_o}$ & $\mathbf{N_{so}}$ & $\mathbf{N_c}$ \\ \hline
        SpaceX Starlink Gen2 C1               & 604                    & 148                        & 144                 & 12            & 12             & 5             \\
        Amazon Kuiper  (Kuiper-V + Kuiper-Ka) & 610                    & 42                         & 2592                & 36            & 72             & 35            \\
        SpaceX Starlink Gen2 C1               & 614                    & 115.7                      & 324                 & 18            & 18             & 1             \\
        Amazon Kuiper  (Kuiper-V + Kuiper-Ka) & 630                    & 51.9                       & 2312                & 34            & 68             & 33            \\
        \end{tabular}%
        }
        \caption{Zonal Example: SpaceX and Amazon have proposed shells that would be separated by a nominal distance of 6 km and 4 km.}
        \label{tab:LowAltitude}
        \end{table*}
        
        \begin{figure}[htbp]
            \centering
            \includegraphics[width=0.65\textwidth]{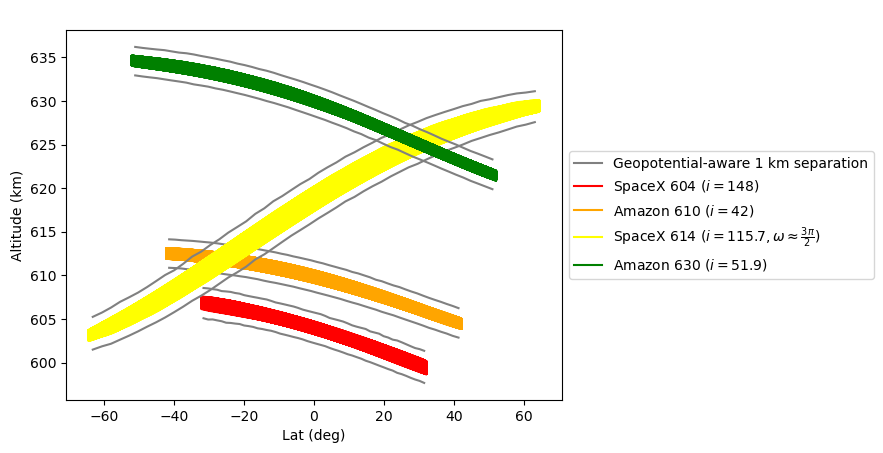}
            \caption{SpaceX and Amazon shells visualized in the latitude-altitude space. The orbit found by the un-constrained GA for the 614 km shell is counter-aligned to the other two shells. (30 day propagation, full force model)}
            \label{fig:LowLEOShells}
        \end{figure}  
        
        \begin{figure}[htbp]
            \centering
            \includegraphics[width=0.65\textwidth]{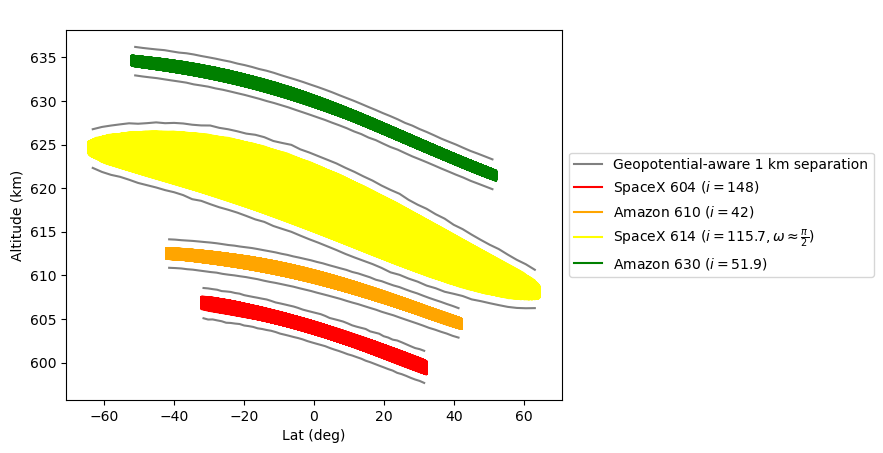}
            \caption{SpaceX and Amazon shells visualized in the latitude-altitude space.  (30 day propagation, full force model)}
            \label{fig:LowLEOShellsAlligned}
        \end{figure}  
    
    \subsection{Example: Repeating Ground Track Method, High Altitude}
    
    In this example, the techniques described in this paper are applied to the CASC shells proposed for a nominal altitude of 1145 km and the Hughes shell proposed at 1150 km. RGT orbits are found at admissible altitudes to stack these 2D-LFCs into five concentric shells.  Each shell is a RGT orbit, seeded with a zonally frozen seed orbit.  While these orbits are sufficient to avoid inter-shell conjunctions, phasing between orbital planes would need to be modified via a different $N_c$ to eliminate intra-shell conjunctions at orbital crossings. Hughes' FCC filing indicates that "[t]o avoid collisions within the constellation, Hughes will make use of a moderate amount of altitude variation such that safe radial separations are maintained at plane intersections."\footnote{The technical appendix to Hughes FCC filing's is available at \url{https://licensing.fcc.gov/myibfs/download.do?attachment_key=13337311}.} The fact that CASC's proposal involves similar overlaps at orbital plane intersections implies that they will likely employ a similar method.  While separating orbital planes in altitude does ensure safety, it does so at the cost of reduced orbital density as compared to a properly phased and maintained shell. The CASC proposal features multiple co-altitude shells with heterogeneous inclinations. In this example, no effort is made to control for the resulting differential RAAN precession.
    
\begin{figure}[ht]
    \centering
    \includegraphics[width=0.5\textwidth]{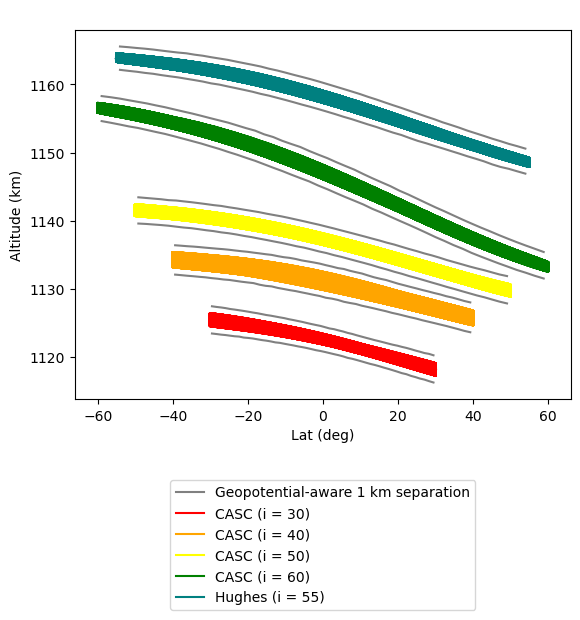}
    \caption{Hughes and CASC inclination-separated shells, visualized in the latitude-altitude space using RGT orbits seeded with GA-generated zonal-frozen orbits (60 day propagation, full force model)}
    \label{fig:HighLEOShells}
    \end{figure}

    As seen in Figure \ref{fig:HighLEOShells}, the sequential ordering of CASC's shells allowed for tighter tessellation between shells. Note that improved density could be achieved by nesting Hughes' shell between the CASC $50\deg$ and $60\deg$ shells as compared to the order presented here. Separation distances are picked notionally. If eccentricity-vector restoring maneuvers were performed more frequently than every two months, shells could also be placed closer than demonstrated in this figure. No analysis is conducted concerning the accumulation of differential drift and phasing error between shells. Eventually this error will also force minor restoring maneuvers. In Table \ref{tab:highLEO}, adjusted altitude refers to the arithmetic mean osculating equatorial altitude of the shell.
    
    \begin{table*}[htbp]
    \resizebox{\textwidth}{!}{%
    \begin{tabular}{llllllllll}
    \textbf{Name}     & \textbf{Nom. Altitude (km)} & $\mathbf{N_{d}}$ & $\mathbf{N_{p}}$ & \textbf{Adjusted Altitude (km)} & \textbf{Inclination (deg)} & $\mathbf{N_{sat}}$ & $\mathbf{N_o}$ & $\mathbf{N_{so}}$ & $\mathbf{N_c}$ \\ \hline

    CASC China SatNet & 1145 & 79  & 6 & 1123.205 & 30 & 1728 & 48 & 36 & 46 \\
    CASC China SatNet & 1145 & 408 & 31 & 1131.249 & 40 & 1728 & 48 & 36 & 46 \\
    CASC China SatNet & 1145 & 79 & 6 & 1137.326 & 50 & 1728 & 48 & 36 & 46 \\
    CASC China SatNet & 1145 & 79  & 6 & 1147.181 & 60 & 1728 & 48 & 36 & 46 \\
    Hughes            & 1150 & 105  & 8 & 1158.289 & 55 & 1440 & 36 & 40 & 18
    \end{tabular}%
    }
    \caption{Repeating Ground Track Example: CASC and Hughes have proposed shells that would be separated by a nominal distance of 5 km.}
    \label{tab:highLEO}
    \end{table*}
    
\section{Conclusions}

Certain regions of LEO are oversubscribed with more interested parties seeking to place large constellations than can be accommodated using naive Keplerian spacing. Orbit design that takes advantage of the Earth's geopotential can allow for closer stacking to more efficiently utilize orbital volume. The methods and results contained in this paper demonstrate that the use of frozen orbits can be combined with geometric analysis in the altitude-latitude plane to ensure non-conjunction between shells. Such shells can be designed as 2D-LFCs and therefore take advantage of considerable previous work on 2D-LFC design. The design burden associated with this technique are modest, as many operators will seek to freeze their orbits and maintain constant energy levels across satellites and orbits for their own station-keeping purposes.  

Both capacity per shell and shell thickness influence aggregate orbital capacity. Nevertheless, denser shell stacking will likely be the more critical driver as demand in LEO continues to increase. Previous work has demonstrated methods for estimating shell capacity and that large shell capacities are possible at reasonable orbital separation distances. Nevertheless, two factors constrain this result. First, our work to date limits shells to a single inclination due to concerns around differential RAAN precession. Second, while sharing a shell across multiple operators is fundamentally feasible, it has not been demonstrated on-orbit and would require extensive coordination among operators who have little incentive to do so if they are not required to do so by congestion or a mandate.

As orbit demand increases, conversations between planned orbital neighbors will become increasingly important to ensure efficient and compatible orbits. Use of common constellation-agnostic methods and schema for discussing slotting and orbit coordination will be helpful to enable interoperability and prevent unsafe states. Such analysis would require information beyond that disclosed in most regulatory filings, including mean eccentricity and eccentricity vector information, intended station-keeping control box sizes (as opposed to general and overly expansive orbital tolerances), and greater precision in distinguishing between nominal, mean, and osculating numbers in orbit disclosures. If operators or regulators wish to adopt these or similar techniques, such sharing and explicit coordination can improve the efficiency with which operators make use of orbital volume. 

There are several areas where this work could be further expanded.  These include:
\begin{enumerate}
\item Further research into the relationship between shell width/slot size and station-keeping maneuver frequency, propulsion type and achievable state accuracy/precision;
\item The use of optimal control theory to generate perfectly closed and periodic reference trajectories through slight modifications to the frozen trajectories developed using the methods in this paper; and
\item Efforts to quantify the gains to capacity and risk reduction from orbital shell separation and slotting.
\end{enumerate}

\section{Acknowledgments}

The Orekit space dynamics library was used to perform propagation, coordinate transforms, and other functions for this work. Miles Lifson was partially supported by the MathWorks Fellowship and the MIT Portugal Partnership 2030 (MPP2030). This work is sponsored in part by the Defense Advanced Research Projects Agency (Grant N66001-20-1-4028). The content of the information does not necessarily reflect the position or the policy of the Government.  No official endorsement should be inferred. Distribution statement A: Approved for public release; distribution is unlimited. The authors want to acknowledge the support of this work by the Air Force’s Office of Scientific Research under Contract Number FA9550-18-1-0115. The authors acknowledge the MIT SuperCloud and Lincoln Laboratory Supercomputing Center for providing high performance computing resources and consultation that have contributed to the research results reported within this paper.

\section{Appendix}
The following are the derivative terms used for the mean elements-based shell width estimation model which include the $J_2$ short-periodic variations. Similar to the corresponding section of the paper, the orbital elements in these formula (but not osculating radius) refer to mean rather than osculating elements.

\begin{equation}
\begin{split}
\frac{dr}{da}&=  \frac{J_{2}\,{\mathrm{R_\oplus}}^2\,\left(\cos\left(2\,\theta +2\,\omega\right)\,{\sin\left(\mathrm{i}\right)}^2-\left(3\,{\cos\left(\mathrm{i}\right)}^2-1\right)\,\left(\frac{2\,\sqrt{1-e^2}}{{\left(e\,\cos\left(\theta \right)+1\right)}^2}+\frac{e\,\cos\left(\theta \right)}{\sqrt{1-e^2}+1}+1\right)\right)}{4\,a^2\,\left(e^2-1\right)}-\frac{e^2-1}{e\,\cos\left(\theta \right)+1}
\end{split}
\end{equation}

\begin{equation}
\begin{split}
\frac{dr}{de}&=\frac{a\,\cos\left(\theta \right)\,\left(e^2-1\right)}{{\left(e\,\cos\left(\theta \right)+1\right)}^2}-\frac{2\,a\,e}{e\,\cos\left(\theta \right)+1}\\
&+\frac{J_{2}\,e\,{\mathrm{R_\oplus}}^2\,\left(\cos\left(2\,\theta +2\,\omega\right)\,{\sin\left(\mathrm{i}\right)}^2-\left(3\,{\cos\left(\mathrm{i}\right)}^2-1\right)\,\left(\frac{2\,\sqrt{1-e^2}}{{\left(e\,\cos\left(\theta \right)+1\right)}^2}+\frac{e\,\cos\left(\theta \right)}{\sqrt{1-e^2}+1}+1\right)\right)}{2\,a\,{\left(e^2-1\right)}^2}\\
&+\frac{J_{2}\,{\mathrm{R_\oplus}}^2\,\left(3\,{\cos\left(\mathrm{i}\right)}^2-1\right)\,\left(\frac{\cos\left(\theta \right)}{\sqrt{1-e^2}+1}-\frac{2\,e}{\sqrt{1-e^2}\,{\left(e\,\cos\left(\theta \right)+1\right)}^2}-\frac{4\,\cos\left(\theta \right)\,\sqrt{1-e^2}}{{\left(e\,\cos\left(\theta \right)+1\right)}^3}+\frac{e^2\,\cos\left(\theta \right)}{\sqrt{1-e^2}\,{\left(\sqrt{1-e^2}+1\right)}^2}\right)}{4\,a\,\left(e^2-1\right)}
\end{split}
\end{equation}

\begin{equation}
\begin{split}
\frac{dr}{d\omega}&=\frac{a\,e\,\sin\left(\theta \right)\,\left(e^2-1\right)}{{\left(e\,\cos\left(\theta \right)+1\right)}^2}-\frac{J_{2}\,{\mathrm{R_\oplus}}^2\,\left(2\,\sin\left(2\,\theta +2\,w\right)\,{\sin\left(i\right)}^2-\left(\frac{e\,\sin\left(\theta \right)}{\sqrt{1-e^2}+1}-\frac{4\,e\,\sin\left(\theta \right)\,\sqrt{1-e^2}}{{\left(e\,\cos\left(\theta \right)+1\right)}^3}\right)\,\left(3\,{\cos\left(i\right)}^2-1\right)\right)}{4\,a\,\left(e^2-1\right)}
\end{split}
\end{equation}

\begin{equation}
\begin{split}
\frac{dr}{di}&=-\frac{\cos\left(\mathrm{i}\right)\,\sin\left(\phi \right)\,\left(\frac{J_{2}\,{\mathrm{R_\oplus}}^2\,\left(2\,\sin\left(2\,\theta +2\,\omega\right)\,{\sin\left(\mathrm{i}\right)}^2-\left(\frac{e\,\sin\left(\theta \right)}{\sqrt{1-e^2}+1}-\frac{4\,e\,\sin\left(\theta \right)\,\sqrt{1-e^2}}{{\left(e\,\cos\left(\theta \right)+1\right)}^3}\right)\,\left(3\,{\cos\left(\mathrm{i}\right)}^2-1\right)\right)}{4\,a\,\left(e^2-1\right)}-\frac{a\,e\,\sin\left(\theta \right)\,\left(e^2-1\right)}{{\left(e\,\cos\left(\theta \right)+1\right)}^2}\right)}{{\sin\left(\mathrm{i}\right)}^2\,\sqrt{1-\frac{{\sin\left(\phi \right)}^2}{{\sin\left(\mathrm{i}\right)}^2}}}
\end{split}
\end{equation}

\bibliography{references2}{}

\end{document}